\documentclass[sigconf]{acmart}
\usepackage{graphicx}
\usepackage[noabbrev,capitalize]{cleveref}
\usepackage{xspace}
\usepackage{enumitem}
\usepackage{subcaption}

\AtBeginDocument{%
  }

\copyrightyear{2026}
\acmYear{2026}
\setcopyright{cc}
\setcctype{by}
\acmConference[CHI '26]{Proceedings of the 2026 CHI Conference on Human Factors in Computing Systems}{April 13--17, 2026}{Barcelona, Spain}
\acmBooktitle{Proceedings of the 2026 CHI Conference on Human Factors in Computing Systems (CHI '26), April 13--17, 2026, Barcelona, Spain}
\acmPrice{}
\acmDOI{10.1145/3772318.3790711}
\acmISBN{979-8-4007-2278-3/2026/04}

\acmSubmissionID{9706}

\newcommand{\co}{CO$_{2}$ }
\newcommand{\baseline}{\textit{No EFSO}}
\newcommand{\mincom}{\textit{EFSO - Minimal}}
\newcommand{\coeq}{\textit{EFSO - \co Equivalency}}
\newcommand{\gamified}{\textit{EFSO - Gamified}}
\newcommand{\social}{\textit{EFSO - Social}}

\begin{document}

\title{Investigating the Effects of Eco-Friendly Service Options on Rebound Behavior in Ride-Hailing}

\author{Albin Zeqiri}
\email{albin.zeqiri@uni-ulm.de}
\orcid{0000-0001-6516-3810}
\affiliation{%
  \institution{Institute of Media Informatics, Ulm University}
  \city{Ulm}
  \country{Germany}
}

\author{Michael Rietzler}
\email{michael.rietzler@uni-ulm.de}
\orcid{0000-0003-2599-8308}
\affiliation{%
  \institution{Institute of Media Informatics, Ulm University}
  \city{Ulm}
  \country{Germany}
}

\author{Enrico Rukzio}
\email{enrico.rukzio@uni-ulm.de}
\orcid{0000-0002-4213-2226}
\affiliation{%
  \institution{Institute of Media Informatics, Ulm University}
  \city{Ulm}
  \country{Germany}
}

\renewcommand{\shortauthors}{Zeqiri et al.}

\begin{abstract} 
Eco-friendly service options (EFSOs) aim to reduce personal carbon emissions, yet their eco-friendly framing may permit increased consumption, weakening their intended impact. Such rebound effects remain underexamined in HCI, including how common eco-feedback approaches shape them. We investigate this in an online within-subjects experiment ($N=75$) in a ride-hailing context. Participants completed 10 trials for five conditions (\textit{No EFSO, EFSO - Minimal, EFSO - \co Equivalency, EFSO - Gamified, EFSO - Social}), yielding 50 choices between walking and ride-hailing for trips ranging from $0.5mi$–$2.0mi$ ($\approx0.80km$-$3.22km$). We measured how different EFSO variants affected ride-hailing uptake relative to a \textit{No EFSO} baseline. EFSOs lacking explicit eco-feedback metrics increased ride-hailing uptake, and qualitative responses indicate that EFSOs can make convenience-driven choices more permissible. We conclude with implications for designing EFSOs that begin to take rebound effects into account.
\end{abstract}


\begin{teaserfigure}
  \includegraphics[width=\textwidth]{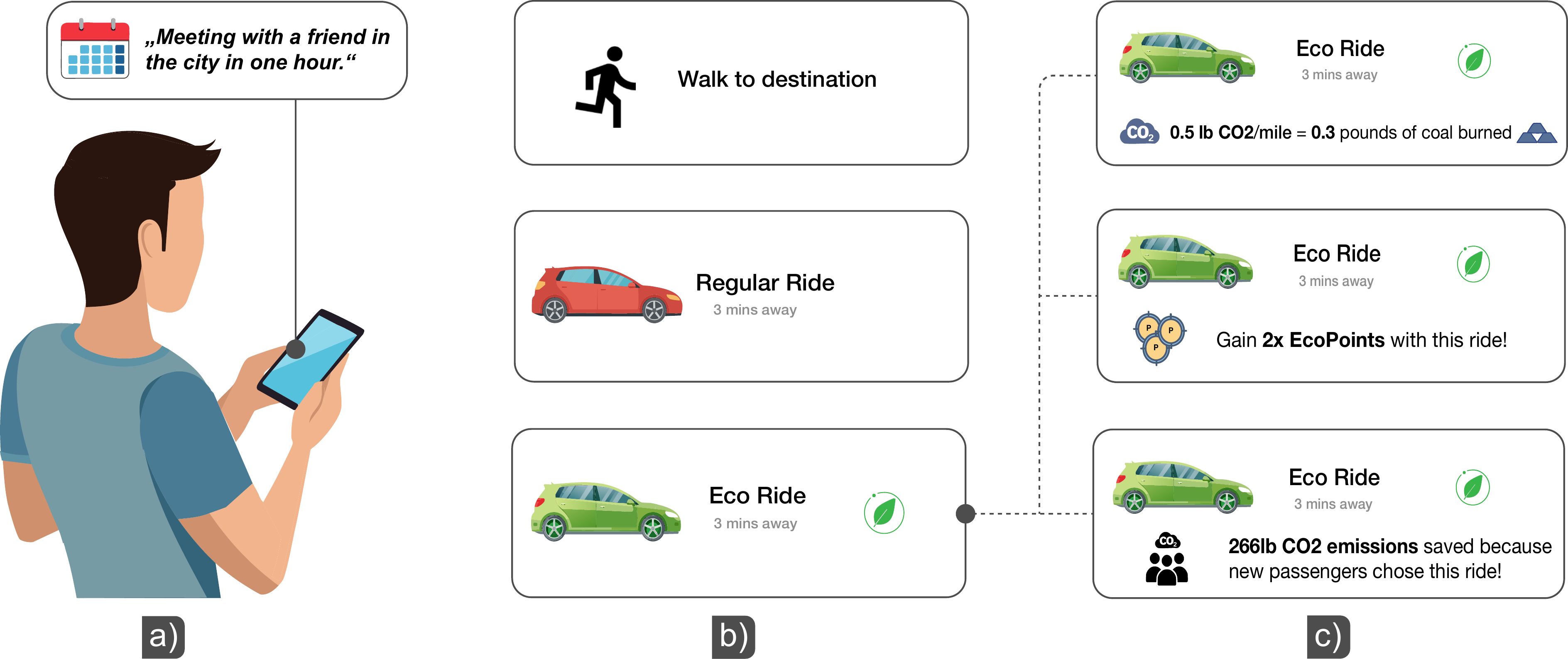}
  \caption{Using an online within-subjects experiment, our work examines the extent to which the introduction of eco-friendly service options (EFSOs) can elicit behavioral responses consistent with rebound effects. Participants imagined traveling to a meeting with a friend (a) in a dense urban environment and chose between walking and a ride-hailing service that either offered a control condition with \baseline~or included an EFSO  that communicated its environmental impact through different metrics ranging from a leaf icon (b) to \co equivalencies, social comparisons, or points (c).}
  \Description{The figure shows the study setup used in the online behavioral experiment. It illustrates a scenario where participants imagine traveling to meet a friend in an urban environment in (a). The following panel (b) shows that participants could choose between walking or using a ride-hailing service that offered either did not offer an eco-friendly option at all (baseline) or offered an eco-friendly ride option that communicated its impact using different metrics (only by a green leaf icon or as shown in panel using carbon dioxide equivalencies, social comparison messages, and point-based feedback).}
  \label{fig:teaser}
\end{teaserfigure}

\maketitle

\section{Introduction}\label{sec:intro}
As anthropogenic climate change intensifies, reducing individual carbon footprints has received growing societal attention~\cite{cantillo_determinants_2025}. This increasingly prompts individuals to reflect on their behavior and adopt pro-environmental practices. Opportunities for personal behavioral changes extend across various domains, including household-focused measures surrounding energy efficiency~\cite{zeqiri2024,bremerReview2022} and waste disposal~\cite{froehlich2010}. Moreover, choices on online platforms for shopping, personal transportation, and travel booking also impact one's carbon footprint. Commonly, today's online platforms offer eco-friendly service options (EFSOs) intended to reduce or offset emissions, such as low-emission ride-hailing choices (e.g., Uber Green\footnote{https://www.uber.com/ride/ubergreen/ - Accessed: 24/01/2026}
), eco-friendly delivery options (e.g., DHL Green\footnote{https://www.dhl.com/us-en/home/express/products-and-solutions/gogreen-plus.html - Accessed: 24/01/2026}
), or carbon offsets on travel booking sites (e.g., Booking.com\footnote{https://sustainability.booking.com/climate-action - Accessed: 24/01/2026}
). Prior studies in Human-Computer Interaction (HCI) examined how to increase the uptake of EFSOs (e.g.,~\cite{laundry2023, mohanty2023}), often leveraging the persuasive qualities of eco-feedback interventions that make their impact salient, for instance, using \co equivalencies, social proof, or gamification.\\

While EFSOs offer a simple and accessible way to reduce personal carbon emissions, the behavioral responses they may elicit remain insufficiently understood. Research in HCI~\cite{bremerRebound2024, bornesReboundSystemic2024}, energy policy~\cite{greening2000}, and ecology~\cite{andrew2024multidisciplinary} point to the fact that efficiency measures are often accompanied by rebound effects that diminish or negate intended environmental benefits. For example, users may purchase an energy-efficient appliance with conservation in mind but then operate it more frequently due to lower operating costs (direct rebound~\cite{sorrell_empirical_2009}), or they may redirect monetary savings toward other resource-intensive actions (indirect rebound~\cite{coroamua2019digital}). Beyond economic mechanisms, recent publications (e.g.,~\cite{reimers2022moral, dutschke_moral_2018, sorrell_limits_2020, santarius_investigating_2016, santarius_how_2018_psychological_mechs}) highlight the relevance of moral-psychological drivers as potential explanatory pathways of rebound effects. In particular, as pro-environmental behavior is often perceived as morally desirable~\cite{urban_pro-environmental_2023, bergquist_most_2020},~\citet{reimers2022moral} suggest that perceived moral credit~\cite{merritt2010moral} gained through a pro-environmental choice can lead individuals to justify less environmentally desirable behavior.\\ 

As such, rebound behavior triggered through moral-psychological pathways can be considered dependent upon subjective perceptions of pro-environmental actions. In the case of EFSOs, this perception could be shaped not only by their availability but also by how their impact is communicated, for instance, through eco-feedback. EFSOs may therefore facilitate behavioral patterns consistent with rebound effects, while eco-feedback mechanisms could further amplify or mitigate these effects by altering perceived impact. Given the lack of consideration rebound effects have received in HCI~\cite{bremerRebound2024}, it is unclear whether EFSOs in digital platforms could elicit behavioral shifts consistent with rebound effects and how eco-feedback shapes this dynamic. This leaves researchers and practitioners with limited guidance on how to design EFSOs that do not undermine their environmental intent.\\

To address this gap, we conducted a within-subjects online experiment ($N=75$) situated in the context of urban transportation (see~\autoref{fig:teaser}). Ride-hailing was chosen as the applied domain because it represents a growing part of the public transportation ecosystem and has been associated with environmental impacts due to increased congestion and vehicle miles traveled~\cite{sheldon2024impacts}. At the same time, recent reviews highlight that electrification and pooling could reduce emissions in this sector~\cite{jenn_emissions_2020}. If, however, the introduction of EFSOs triggers rebound effects that offset these potential gains, their overall environmental benefit may be compromised, making ride-hailing a relevant target context for investigating how EFSOs influence user behavior. 
In our study, participants were asked to imagine themselves in an urban travel scenario in which they chose between walking and using ride-hailing for trips at varying walkable distances according to previous work ($0.5mi - 2.0mi$~($\approx0.8km-3.2km$)~\cite{watson_walking_2015}. As a potential case of behavioral patterns consistent with direct rebound, we examined how the introduction of EFSOs (~\autoref{fig:teaser}b), combined with different eco-feedback concepts (\coeq, \social, and \gamified~\autoref{fig:teaser}c), influenced the likelihood of choosing ride-hailing over walking, compared to a control without EFSOs (\baseline). Additionally, we collected reflections and open-ended feedback in which participants described their rationales when switching from walking to ride-hailing and their thoughts on EFSOs more broadly. This was done to gain additional insights into decision-making processes in the presence of EFSOs.\\

We found that the introduction of an EFSO significantly increased the likelihood of ride-hailing uptake when the EFSO was visualized using a simple leaf icon as feedback (\mincom). We also found an increased likelihood of ride-hailing uptake for EFSOs that used eco-feedback (\coeq, \social), specifically for trips with long walking distances $1.35mi-2.0mi$~($\approx2.17km-3.22km$). However, not all eco-feedback variants lead to detectable shifts in ride-hailing uptake (\gamified). Additionally, qualitative responses suggest that participants viewed choosing an EFSO as motivating and generally worthwhile, though not as a strongly impactful pro-environmental action. Decisions within the simulated choice paradigm were primarily convenience-driven. The presence of an EFSO became salient only once walking was perceived as sufficiently effortful, at which point participants started considering EFSOs as a justification for switching from walking to ride-hailing. These insights provide support for the need to carefully consider how EFSOs and eco-feedback are integrated into existing services to avoid undermining their intended environmental benefits. In sum, our work contributes the following:
\begin{enumerate}
\item We contribute empirical insights from an online within-subjects experiment ($N=75$) in the ride-hailing context that demonstrate how the introduction of EFSOs can facilitate behavioral patterns consistent with direct rebound. Our findings further show how common eco-feedback concepts influence users’ decisions and perceptions of environmental impact. 
\item We propose a set of practical implications for the design and implementation of EFSOs, highlighting how design choices can inadvertently foster or mitigate rebound effects and informing more ecologically responsible user interface design. In doing so, we argue that our work responds to recent calls by the Sustainable HCI (SHCI) community to examine rebound effects more directly, while also extending prior research in the automotive domain that has primarily focused on the persuasive qualities of eco-feedback.
\end{enumerate}

\section{Background \& Related Work}\label{sec:rel_work}
Our work builds on previous research on interventions for pro-environmental behavioral change, rebound effects, and moral balancing as a possible theoretical framework underlying rebound effects.
\subsection{Rebound Effects}\label{subsec:rel_rebounds}
\subsubsection{Definition \& Typology}\label{subsubsec:typology-rebound}
In energy literature, rebound effects are defined as the difference in potential environmental benefits (PEB) of an intervention and the actually achieved environmental benefits (AEB)~\cite{vivanco2016}. In absolute terms, rebound effects can be expressed as $RE=PEB-AEB$ and as $\%RE=(\frac{PEB-AEB}{|PEB|})x100$ in relative terms~\cite{vivanco2014}. As such, rebound effects can occur as a diminishing (i.e., $RE<100\%$), neutralizing (i.e., $RE=100\%$), or even negating effect of the initial energy saving potential (i.e., $RE>100\%$)~\cite{vivanco2016}. Prior work describes different types of rebound effects. \textit{Direct rebound effects} occur when efficiency improvements in a domain lead to increased consumption in the same domain~\cite{sorrell_empirical_2009, bremerRebound2024, coroamua2019digital}, for instance, when the energy efficiency makes using the same service or product more affordable. \textit{Indirect rebound effects} can occur due to different mechanisms. For instance, when efficiency gains free up resources such as money, this can lead people to spend savings on other goods or services~\cite{berkhout_defining_2000} that also consume resources (income effect~\cite{incomeEffect}). Similarly, improvements in energy efficiency can reduce the relative cost of a product, leading consumers to substitute it for more expensive alternatives and thereby increase its utilization (substitution effect). \textit{Economy-wide} or \textit{structural} rebound effects additionally include systemic shifts in markets, production, and consumption that collectively influence environmental outcomes \cite{greening2000, jevonsSorrell2009}.

\subsubsection{Underlying Mechanisms}
Most commonly, economic mechanisms (i.e., affordability, reallocation of monetary spending) are positioned as typical drivers of rebound effects~\cite{reimers2022moral, berkhout_defining_2000}. While economic explanations, such as the reinvestment of financial savings, are well-documented, moral-psychological mechanisms have also been highlighted as predictors~\cite{reimers2022moral, dutschke_moral_2018, peters2016exploring, santarius_how_2018_psychological_mechs, santarius_investigating_2016}. Explanations for rebound effects that point to moral-psychological mechanisms are frequently grounded in moral balancing theory, which proposes that individuals evaluate their behaviors against one another to maintain a balanced moral self-image~\cite{merritt2010moral, ploner2013self}. Moral balancing encompasses both moral licensing and moral cleansing~\cite{lee2013ACM}. Moral licensing occurs when previous ``good'' actions are used to justify subsequent behaviors that are less morally desirable~\cite{lee2013ACM, merritt2010moral, blanken2015meta}. In contrast, moral cleansing refers to the tendency to compensate for an undesirable action by subsequently engaging in behavior that is perceived as virtuous. One explanation for this mechanism suggests that individuals \textit{accumulate} moral credits from prior desirable actions, which can later be \textit{spent} to justify less ethical behaviors~\cite{merritt2010moral, hollander1958conformity}. The moral credentials model suggests that after engaging in a morally positive action, individuals feel they have demonstrated their moral character, allowing them to engage in behaviors that might otherwise be perceived negatively without harming their self-image~\cite{merritt2010moral}. Building on moral balancing theory, \citet{reimers2022moral} propose the rebound cube, a framework that categorizes rebound effects by the underlying mechanism, distinguishing between economic and moral-psychological causes. Economic direct rebound occurs when people use the same service or product more frequently because efficiency improvements make it more affordable. In contrast, moral-based direct rebound~\cite{reimers2022moral} is driven by perceived moral credit rather than monetary savings. For example, while someone may previously have avoided ordering takeout to reduce packaging waste, they may now feel justified in doing so regularly when the platform offers a green delivery option, reasoning that the eco-friendly service offsets the impact (i.e., reduced loss of moral credit). This logic also extends to other types of rebound as well (see e.g.,~\cite{reimers2022moral, dutschke_moral_2018}). The described effects are inherently subjective, as they depend on whether individuals attribute significance to a pro-environmental action in the first place, which in turn is shaped by factors such as pro-environmental attitudes~\cite{gholamzadehmir_moral_2019}.

\subsubsection{Sustainability \& Rebound Effects in HCI} 
Prior HCI research in the Sustainable Human-Computer Interaction (SHCI) community has long focused on developing and evaluating persuasive approaches to encourage individual behavioral change~\cite{disalvoMapping2010, bremerReview2022}. Commonly studied use cases included residential resource conservation and waste management~\cite{disalvoMapping2010, silberman2014next, persuasionUnpersuaded2012, bremerReview2022}, as well as transportation~\cite{colleyEco2022, park2017ecotrips, sanguiettyFlight2022}. Interventions across these contexts commonly highlight eco-friendly options (e.g.,~\cite{laundry2023}) or provide eco-feedback to raise awareness of environmental consequences connected to a behavior to then make a sustainable choice~\cite{sanguinetti2018}. Previous work in this area (e.g.,~\cite{persuasionUnpersuaded2012, adaji2022review, mencariniInteractions}) has emphasized that interventions targeting individual behavior change are inherently limited in effectiveness because they focus on the individual user. Achieving lasting behavioral change is complex, as it requires alignment with everyday routines~\cite{ganglbauer2013activist} and must account for the numerous contextual factors that shape decision-making~\cite{zeqiri2024, van2022select}. Furthermore, the true impact of behavior change interventions is poorly understood, as investigation of rebound effects is often not a primary focus~\cite{bremerRebound2024}. \citet{bremerRebound2024} conducted a literature analysis examining how rebound effects have been considered in HCI, with a focus on the smart home context. They similarly found that rebound effects have rarely been an explicit focus in HCI research. \citet{bremerRebound2024} consequently point out the need to better understand ways to identify, measure, explain, and mitigate causal rebound effects caused by conservation interventions. In this study, we examine, specifically, how EFSOs may trigger behaviors that lead to rebound effects. While the use of EFSO focuses on individual behavior, the scale at which these service options are presented to individuals, and the environmental implications of ride-hailing~\cite{sheldon2024impacts}, which we use as an applied context in this work, represent a relevant target for understanding behavioral responses that lead to rebound effects.

\subsection{Eco-Friendly Service Options in Online Platforms}\label{subsec:efso-general-rel}
\subsubsection{Environmental Impacts of Online Actions}\label{subsub:environmental-actions-rel}
As essential services offer online availability, environmentally relevant decisions increasingly occur through digital platforms. Interfaces such as dashboards, control panels, and menus mediate eco-friendly choices (e.g., transportation options~\cite{mohanty2023, sanguiettyFlight2022, park2017ecotrips}). Prior work indicates that, at scale, the cumulative effect of decisions in online contexts can be substantial~\cite{belkhir_assessing_2018, zulfiqar_digitalized_2023}. This applies both to services that are purely digital (e.g., cloud storage, streaming, messaging) and to services where the platform acts as an intermediary for physical outcomes (e.g., online shopping, travel booking, or personal transportation). For instance, \citet{bremerDataCenters2025} reports that user behavior in cloud storage services can account for up to 26\% of a data center’s energy consumption~\cite{dataCenterSurvey2016}, while \citet{streamingCarbon2019} show that behavioral patterns in streaming services drive rising resource demand. Given the scale at which consumption occurs through digital platforms, shifts in user behavior can accumulate into notable environmental impacts. This makes digital platforms a critical context for integrating EFSOs, as their design and framing may determine whether large populations make choices that align with or undermine environmental goals.

\subsubsection{Promoting Eco-Friendly Service Options}\label{subsub:efsos-rel}
EFSOs are service options that allow users to select alternatives with lower environmental impact. Examples include sustainable delivery choices offered by logistics providers (e.g., FedEx\footnote{https://www.fedex.com/en-us/sustainability.html - Accessed: 24/01/2026}, DHL\footnote{https://www.dhl.com/us-en/home/express/products-and-solutions/gogreen-plus.html - Accessed: 24/01/2026}), carbon offset programs on travel booking sites (e.g., Booking.com\footnote{https://sustainability.booking.com/climate-action - Accessed: 24/01/2026}
), and low-emission ride categories in ride-hailing services (e.g., Uber Green\footnote{https://www.uber.com/ch/de/ride/ubergreen/ - Accessed: 24/01/2026}
). While implementations vary across domains, the goal of EFSOs is to make sustainable choices visible, accessible, and simple to adopt.
HCI research has extensively investigated how to encourage pro-environmental decision-making, both through mechanisms such as EFSOs on digital platforms and in broader everyday contexts, including household energy use, water consumption, and recycling. While EFSOs represent one explicit and increasingly common form of promoting sustainable behavior, they fit into a wider body of work concerned with how interaction design can support environmentally responsible choices in daily life. Experimental studies in recent work suggest that introducing eco-feedback interventions can promote eco-friendly choices~\cite{laundry2023, mohanty2023}.~\citet{froehlich2010} (building on~\citet{holmes2007eco} and~\citet{mccalley1998computer}), define eco-feedback as ``\textit{[...]technology that provides feedback on individual or group behaviors with a goal of reducing environmental impact.''}~\cite[p.1]{froehlich2010}. Examples of such interventions include approaches that make energy use tangible (e.g., via virtual or augmented reality applications~\cite{preussnerVREcoFeedback, mohanty2023}),~\textit{social proof} mechanisms that highlight the behavior of peers~\cite{socialBollinger2023, mohanty2023, colleyEco2022}, modified choice architectures such as defaults or nudges~\cite{laundry2023}, and gamified elements such as (consumption) points~\cite{mohanty2023, laundry2023} or badges for sustainable actions~\cite{andersonBadges, mulcahy_game_2021}. To clarify the dimensions relevant to the design of eco-feedback concepts,~\citet{sanguinetti2018} proposed a conceptual framework based on a review of literature surrounding eco-feedback design across different academic fields. They uncover three main dimensions of feedback timing (i.e., when feedback is presented), information (i.e., what information is presented), and display (i.e., how and where information is presented)~\cite{sanguinetti2018}. Building on this framework,~\citet{mohanty2023} examined how different carbon representations and framings of carbon emissions can influence users to choose eco-friendly ride-hailing options. For instance, they reported that compared to a minimal control condition that shows only a leaf icon to indicate an eco-friendly mode, providing additional eco-feedback in the form of~\textit{raw \co amounts with equivalencies} (e.g., visualizing gallons of gasoline consumed equivalent to a ride's carbon emissions) and ~\textit{gamified} approaches that provide digital points, were effective in promoting eco-friendly ride choices in an online experimental study~\cite{mohanty2023}.  

\subsection{Summary \& Research Gap}\label{subsub:gap-rel}
Given the scale of consumption contexts navigated through online platforms, EFSOs represent an accessible way to incorporate pro-environmental choices into one's daily routine to reduce personal carbon footprints. Prior work in HCI has examined how eco-feedback can be used to promote the use of EFSOs (e.g.,~\cite{mohanty2023, laundry2023, sanguiettyFlight2022, visser2023get}). Findings from work such as \citet{mohanty2023} emphasize the benefits of EFSOs by highlighting concepts that convey \textit{raw \co amounts with equivalencies}, \textit{social proof}, and \textit{gamification} (e.g., via points). 

In doing so, eco-feedback directly shapes how impactful EFSOs are perceived to be. Since pro-environmental choices, such as selecting an EFSO, are typically regarded as virtuous~\cite{urban_pro-environmental_2023}, and eco-feedback can amplify perceptions of their impact through its metrics, there is potential for shifts in perceived moral credit to occur. Given existing knowledge that individuals tend to balance trade-offs between their self-perception and their actual behaviors~\cite{festinger1962cognitive}, shifts in perceived moral credit can meaningfully alter how such trade-offs are resolved. In particular, according to~\citet{reimers2022moral}, earning perceived moral credit through a pro-environmental choice can lead individuals to justify less environmentally desirable choices within the same (moral-based direct rebound) or a different domain (moral-based indirect rebound). Given the growing body of work highlighting moral-psychological mechanisms as factors involved in the emergence of rebound behavior (e.g.,~\cite{reimers2022moral, dutschke_moral_2018}), this raises the possibility that EFSOs, together with the eco-feedback mechanisms intended to promote them, may paradoxically limit carbon-saving potential. 

Nevertheless, rebound effects remain underexplored in HCI~\cite{bremerRebound2024, bremerReview2022}. It remains unclear whether EFSOs can trigger behavioral patterns that result in rebound effects and how established eco-feedback concepts shape the extent of these outcomes. It is equally relevant to explore how users rationalize their decision-making processes in the presence of EFSOs, as these reflections reveal underlying motives and justifications that can drive such choices. Without clarity on these aspects, researchers and practitioners are constrained in their ability to design and evaluate EFSOs that reliably achieve their intended environmental benefits.

\section{Methodology}\label{sec:method}
As outlined in~\autoref{subsec:rel_rebounds}, rebound effects vary depending on their underlying drivers and the context in which they occur. Compared to indirect or structural rebound effects, direct rebound effects are better documented in the literature~\cite{sorrell_empirical_2009}. The present study, therefore, concentrates on behavioral responses that are triggered through EFSOs and whether they could lead to direct rebound effects. We investigate this in the applied context of ride-hailing (urban transportation). Although prior work has examined eco-friendly driving and the use of EFSOs in this sector~\cite{mohanty2023, colleyEco2022}, the possibility that EFSOs themselves could induce rebound effects has not been considered. Our experiment was guided by the following research questions (RQs):
\begin{itemize}
    \item [] \textbf{\textit{RQ1}} How does introducing EFSOs shape behavioral responses associated with direct rebound effects in a ride-hailing context?
    \item [] \textbf{\textit{RQ2}} How does eco-feedback in EFSOs that uses \textit{raw \co amounts with equivalencies}, \textit{social proof}, or \textit{gamification} affect whether behavioral responses associated with direct rebound effects arise in a ride-hailing context?
    \item [] \textbf{\textit{RQ3}} How do users rationalize ride-hailing choices when EFSOs are available?
\end{itemize}

RQ1 and RQ2 investigate whether EFSOs and the design mechanisms used to promote them trigger behavioral patterns consistent with direct rebound. RQ3 explores the factors users consider when making decisions in the presence of EFSOs. 

To address these questions, we designed an online choice experiment situated in the ride-hailing context. Participants were asked to imagine themselves in urban travel scenarios where they had to attend a meeting with a friend at a public location and decide whether to reach the destination by walking or using ride-hailing services. Each participant experienced five EFSO variants for trips at various distances in counter-balanced order. A \baseline~variant (i.e., only gas-powered vehicles) and four variants that included an EFSO (electric vehicle). The EFSO was visually denoted with either a simple leaf icon (\mincom), raw \co amounts with equivalencies (\coeq), social proof (\social), or gamification via points (\gamified). For each variant, participants made ten trip decisions with distances between $0.5mi- 2.0mi$ ($0.8km-3.22km$). Based on our review of prior work (see \autoref{sec:rel_work}), we derived the following hypotheses.

\begin{itemize}
    \item [] \textbf{\textit{H1}}: Participants are more likely to choose ride-hailing over walking when EFSOs are present compared to when they are not present.
    \item [] \textbf{\textit{H2}}: Participants are more likely to choose ride-hailing over walking when \co amounts and equivalencies are used compared to a minimal communication condition where the EFSO is highlighted only by a leaf icon.
    \item [] \textbf{\textit{H3}}: When switching from walking to ride-hailing, participants perceive less moral credit loss when EFSOs are present compared to when they are not present.
\end{itemize}

In their work, Bremer et al. highlight that ``\textit{Several studies [...] point to the potential to better understand the causal factors of rebound and to identify and evaluate rebound patterns that can inform research and design}''~\cite[p.14]{bremerRebound2024}. Our RQs and hypotheses (in particular H1 and H2) are directly aimed at investigating a potential causal mechanism between EFSO integration and behavioral responses consistent with direct rebound. Revealing such mechanisms in a context such as decision-making on ride-hailing adoption requires maintaining control over contextual factors (e.g., trip purpose, time pressure, route characteristics) that are inherently interrelated in ride-hailing decisions while systematically manipulating EFSO variants to maintain the necessary internal validity. We employed a behavioral paradigm study in the form of a choice task to test our hypotheses. As noted in a recent review by \citet{lange2023behavioral}, such paradigms are widely used because they offer the necessary control for manipulating variables involved in hypothesized causal mechanisms and contextual influences~\cite{lange2023behavioral, berger2021measuring}. They are also well-suited for integration into online surveys~\cite{lange2023behavioral}, which allows efficient data collection from target demographic groups. In addition, our approach aligns with numerous previous HCI studies that have also employed online behavioral experiments to investigate pro-environmental decision-making (e.g., \cite{mohanty2023, laundry2023, visser2023get}). 
In the following, we introduce the experimental task, the independent variables, measures, and standardized contextual factors of our experimental design. Lastly, we describe the experimental procedure and analysis.

\subsection{Task}\label{subsec:task}
The experimental task was designed to create an interpretable setting for examining how EFSOs influence behavioral responses consistent with direct rebound. The following factors were taken into consideration to derive such a scenario.
One pathway through which the integration of EFSOs in ride-hailing platforms may lead to direct rebound effects is an increased likelihood of choosing ride-hailing for trips where individuals would otherwise have opted for zero-emission transportation alternatives. Such effects are most clearly identifiable when ride-hailing uptake is contrasted against a truly zero-emission mode, such as walking. Prior work indicates that distances up to about 2.0 miles ($\approx 3.22$ km) are commonly perceived as walkable in urban U.S. contexts~\cite{watson_walking_2015, yang2012walking}, which makes walking a suitable benchmark for assessing substitution toward higher-emission travel. In naturalistic mobility settings, individuals can choose among multiple transportation modes (e.g., walking, cycling, public transit, and various ride-hailing options), each associated with distinct trade-offs in cost, time, convenience, and environmental impact. However, incorporating this full multimodal choice set into an experimental design would introduce substantial confounding influences and obscure the specific behavioral effects of EFSOs. Because the feasible distance ranges of cycling or public transit overlap more extensively with ride-hailing than those of walking, rebound-like substitution would also be more difficult to detect.
In our experiment, participants were asked to imagine themselves in the following scenario:\\ 

\textit{``In one hour, you are meeting a friend at a public location in the city, as shown on the map [map visualization on the same page]. The city has good walking infrastructure. For each trip, you will see the route, distance, and estimated duration for both walking and using ride-hailing. Please decide for each trip whether you would use ride-hailing or walk, using the choice interface provided below."}\\

We standardized the scenario's contextual influences by ensuring sufficient time to walk (max. $2.0mi$ $\approx$ 40 minutes, and meeting in one hour), a walkable urban environment (see~\autoref{sec:procedure}), and a leisure-oriented trip purpose framed as meeting a friend in the city.

\begin{figure*}[t]
    \centering
    \includegraphics[width=0.99\linewidth]{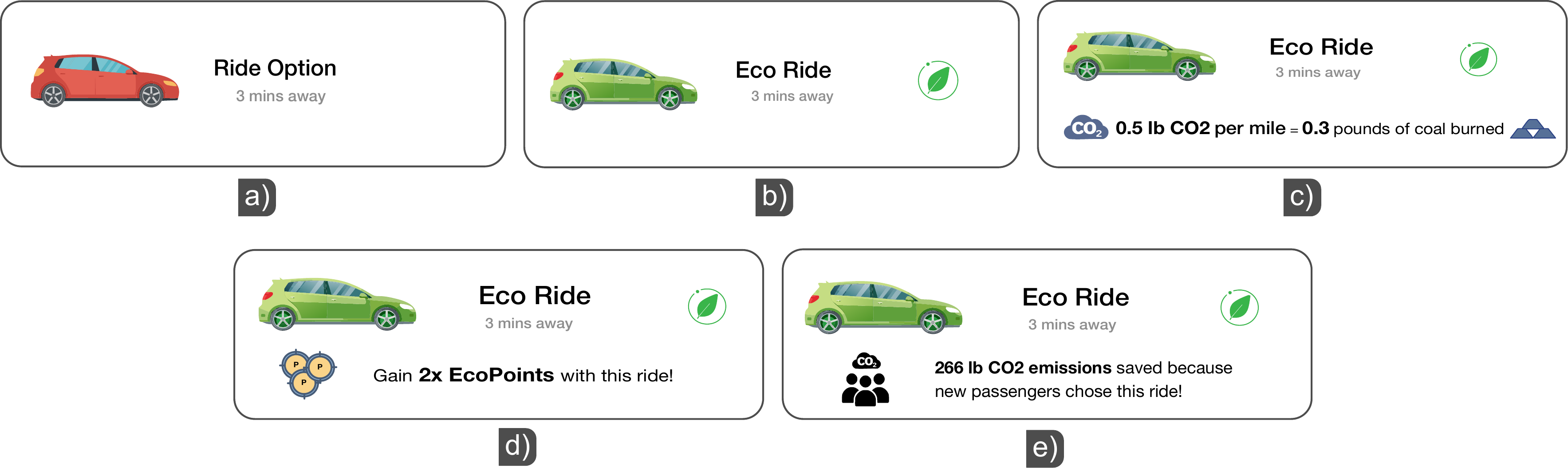}
    \caption{Visualizations of the ride-hailing variants used in our study (derived from~\citet{mohanty2023}). a) represents a regular gas-powered vehicle. In the \baseline~variant, only gas-powered ride-hailing options exist. b)-d) represent eco-friendly ride options. In b), the eco-friendly ride is shown using a green leaf icon and not additional information (\mincom). In c) the raw \co output is communicated together with more relatable carbon equivalencies (\coeq). In d) the positive environmental impact of the eco-friendly ride is represented using a gamified metric based on ``EcoPoints'' (\gamified). In e) the environmental impact of ride options is communicated using the collective \co emissions by users within the last week (\social).}
    \label{fig:designs}
    \Description{The figure shows the different ride-hailing options presented to participants in the study. Panel (a) displays a standard gas-powered ride, which is the only option available in the baseline condition. Panels (b) through (e) show eco-friendly ride options. In panel (b), the eco-friendly ride is marked only with a simple green leaf icon. In panel (c), the ride-hailing options include information about carbon emissions along with everyday equivalencies to help interpret the values. In panel (d), the environmental implications are expressed through a gamified points system called "EcoPoints". In panel (e), the option highlights the collective carbon emissions by all users in the previous week.}
\end{figure*}

\subsection{Independent Variables}\label{subsubsec:independent_vars}
\subsubsection{EFSO Variant}  
Based on our review of the literature on pro-environmental behavior change and eco-feedback design, particularly in ride-hailing contexts, we incorporated the following EFSO variants. \autoref{fig:designs} shows each of the designs used in our study.\\ 

\textit{EFSO - Equivalency:} Visualizing the environmental impacts of a service option through raw \co amounts has been shown to increase awareness and encourage more sustainable choices~\cite{mohanty2023, sanguiettyFlight2022}. In our study, the eco-friendly ride option displayed the expected emission savings in pounds of \co compared to the conventional ride. \citet{mohanty2023} recommend presenting raw \co values as the primary representation and found that this format was the most effective in promoting the selection of EFSOs in their study, even though users rated it as less relatable. To make \co values more relatable and increase perceived usefulness, they further experimented with equivalency-based framings (e.g., pounds of coal burned) that connect raw values to more familiar concepts. We combined both approaches in our design, treating raw \co as the central representation while using equivalencies only as an aid to interpretation. Equivalencies were calculated using the EPA equivalencies calculator~\cite {us_epa_calculator}.\\
 
\textit{EFSO - Social:} 
By leveraging social comparisons, it can be highlighted that others are engaging in a particular behavior. \citet{laundry2023} examined the effect of social comparison in eco-feedback to encourage energy-saving choices during laundry tasks, reporting that such framings effectively increased sustainable behavior. Similarly, \citet{mohanty2023} showed that presenting the impact of eco-friendly ride options as the collective choice of all users was effective in promoting their adoption. Similarly, the \social~variant was accompanied by a message indicating the combined pro-environmental impact of passengers in the past week.\\

\textit{EFSO - Gamified:}  
Gamification incorporates elements such as points or rewards to encourage pro-environmental behavior. For this variant, participants earned more ``Eco-Points'' for selecting the eco-friendly ride, which accumulated across trials to create a sense of progress. \citet{mohanty2023} found that even simple point-based feedback, without further explanation, was effective in promoting the use of eco-friendly ride options. Similar effectiveness of gamified concepts was reported for various other contexts in a literature review by~\citet{douglas2021gamification}. As such, we implemented a gamified variant with a visual display of points that are awarded upon selection of the ride-hailing option (see~\autoref{fig:designs}d).\\  

\textit{EFSO - Minimal:}
To investigate the influence of the specific eco-feedback designs, we added a variant in which an eco-friendly ride was available but only neutrally highlighted with a green leaf icon to indicate its positive environmental impact, similar to the visualization used by~\citet{mohanty2023} (see \autoref{fig:designs}b).\\   

\textit{No EFSO}:  
Based on RQ1 and RQ2, we examine whether the introduction of EFSOs increases the likelihood of rebound effects and how the different eco-feedback concepts shape their severity. To investigate the general effect of EFSOs, we included a control variant in which no eco-friendly ride option was available.\\

The \gamified, \social, and \coeq~ variants differ in the metrics used to communicate environmental impact. Metric representation is a key design dimension of eco-feedback~\cite{sanguinetti2018}. We focused on this dimension because \citet{mohanty2023} demonstrated in their study that different carbon equivalencies influence how users interpret and assess emissions information in ride-sharing contexts, thereby affecting the perceived usefulness and relatability of eco-information. In their second study, \citet{mohanty2023} examined framings that varied in ambiguity, social emphasis, and valence. Although these framings showed persuasive effects, they primarily influenced how the information was interpreted motivationally (e.g., by highlighting positive versus negative outcomes) rather than altering the underlying informational content. Because our aim was to examine whether differences in how environmental impact is understood affect ride-hailing choices and shape perceptions of one’s personal impact, varying the metric dimension enabled us to manipulate the informational content of EFSOs directly.\\

\subsubsection{Distance}
As outlined in~\autoref{subsec:task}, direct rebound in ride-hailing may manifest as an increased ride-hailing likelihood. To capture this effect, \textit{Distance} was included as an independent variable. Prior studies~\cite{yang2012walking, watson_walking_2015} indicate that in urban contexts, trips exceeding $2mi~(\approx3.22km)$ are rarely perceived as walkable. In contrast, trips below $0.5mi~(\approx0.8km)$ are too short to plausibly involve ride-hailing. Accordingly, we selected a distance range of $0.5mi-2.0mi~(0.8km-3.22km)$ miles and divided it into ten increments to achieve consistent variation across the distance range.\\ 

\subsection{Measures}\label{subsec:dependent_vars}
\subsubsection{Rebound Behavior}\label{subsubsec:dep_var_rebound_behavior}
In this study, direct rebound effects are operationalized as an increased likelihood of choosing ride-hailing despite the availability of a zero-emission alternative such as walking. This pattern is particularly relevant for shorter trips, where walking remains a feasible option. A higher likelihood of ride-hailing uptake is thus interpreted as an indication of reduced self-imposed restraint on environmentally costly behavior in favor of values such as comfort and efficiency.

\subsubsection{Moral Credit}\label{subsubsec:dep_var_moral_credit}
To assess perceived shifts in moral credit, we adapted an item from~\citet{lin2016ethical}. However, their question was framed in terms of moral credit gain. In our case, we aimed to measure whether participants perceived moral credit loss, which is why we negated the original item by~\citet{lin2016ethical} and left the remaining formulation unchanged. The resulting single item was rated on a 7-point Likert scale (Strongly disagree=1, Strongly agree=7) and formulated as follows: \textit{``I've lost moral credit through my choices in this trip.''}. To ensure a standardized understanding of the concept of moral credit, the question text included an explanation of the concept of moral accounting and moral credit taken from~\citet{dutschke_moral_2018}. 

\subsubsection{Perceived Environmental Impact}\label{subsubsec:dep_var_perceived_impact}
We included a single item to measure how high participants perceived the environmental burden of the ride-hailing service to be in comparison to walking. This was rated on a 7-point Likert scale (Strongly disagree=1, Strongly agree=7), and the item was formulated as follows: \textit{``Compared to walking, my transportation choice caused significant environmental damage.''}. The formulation aimed to make the reference point of the comparison explicit and to ensure that participants assessed not only the presence of a difference but also the perceived magnitude of that difference.

\subsubsection{Perceptions \& Rationales}\label{subsubsec:dep_var_perceptions}
To complement the above measures and address RQ3, participants were asked to provide free-text reflections describing their decision-making process in as much detail as possible.

\subsection{Control Variables}
As outlined in~\autoref{subsec:task}, trip purpose, time constraints, and route characteristics were standardized through the scenario description to ensure that observed differences in choice behavior could be attributed to the experimental manipulations rather than contextual variability. Because ride-hailing adoption is also influenced by monetary considerations, price was included as an additional control variable. The following section summarizes how each of these aspects was held constant.\\

\subsubsection{Price} Ride-hailing price was calculated by averaging the cost of a two-mile trip with Uber and Lyft across ten major U.S. cities and then calculating proportional prices for each of the ten trip distances. The eco-friendly ride costs one dollar more than the regular ride.
\subsubsection{Location \& Route Infrastructure} No specific city was named, but participants were informed that the walking infrastructure was suitable for walking. In particular, participants were informed that sidewalks were available for the entire route and no heavy traffic would be encountered, to eliminate uncertainty about the feasibility or safety of walking.
\subsubsection{Trip Purpose \& Time Constraints} The purpose of each trip was consistently described as meeting a friend with enough time (one hour) for the trip, even at the longest distance (2.0 miles/40-minute walk).\\

\begin{figure*}[h]
    \centering
    \includegraphics[width=0.9\linewidth]{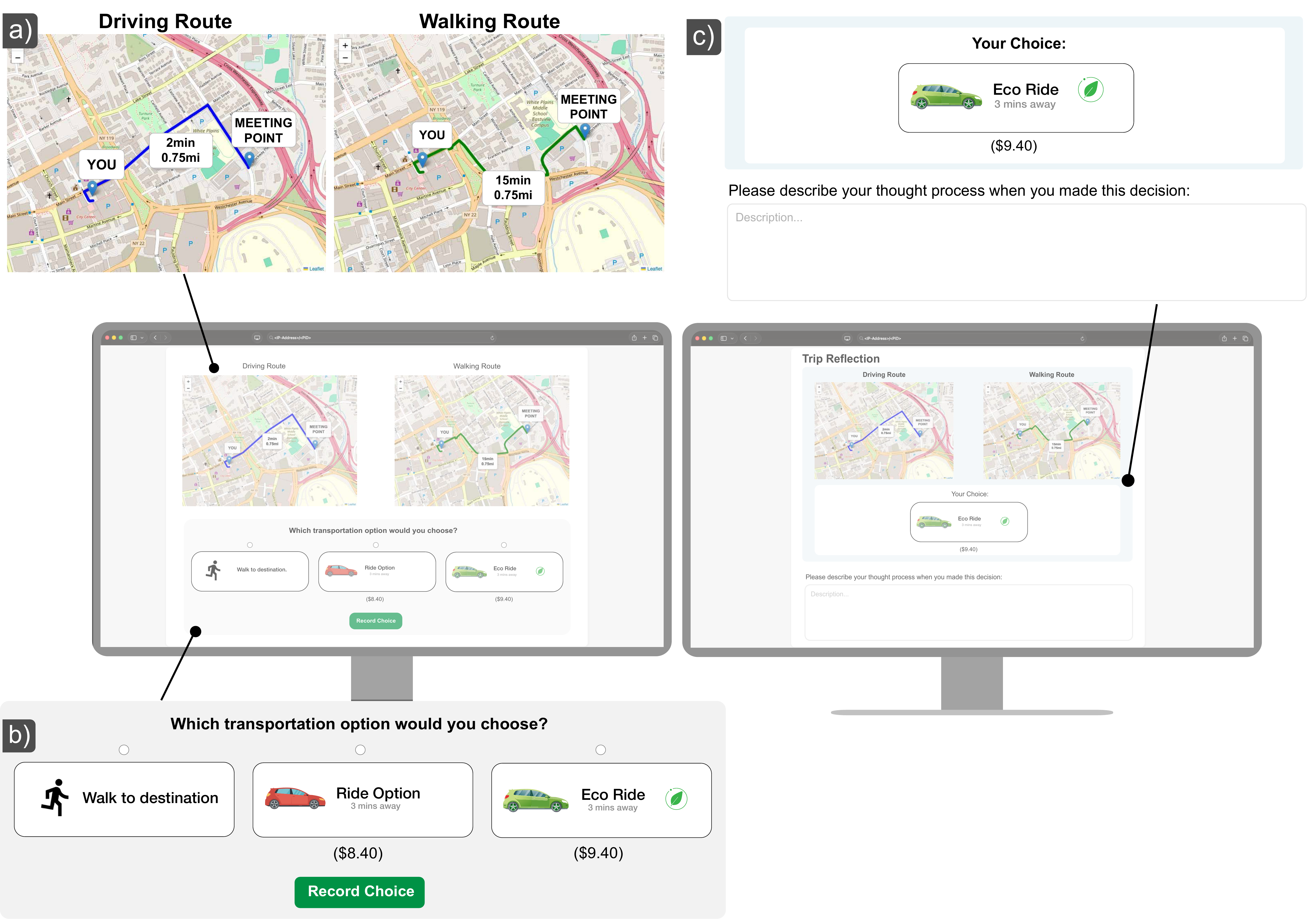}
    \caption{The figure shows the interface of our study application. Panel (a) presents the route information section, which displays walking and driving maps along with the meeting location, starting point, distance, and estimated travel duration. Panel (b) shows the choice interface, featuring two ride-hailing options and a button to select walking. Panel (c) displays the reflection interface, where participants were shown the shortest distance at which they switched from walking to ride-hailing for each EFSO variant and were prompted to provide a free-text rationale for that decision.}
    \label{fig:study-interface}
    \Description{The figure shows three screens from the study’s web application. The first screen displays two small maps: one for walking and one for driving, along with the starting point, meeting location, distance, and estimated travel time. The second screen presents the decision interface, offering two ride-hailing options and a separate button to choose walking. The third screen shows a reflection prompt where participants are told the shortest distance at which they switched from walking to using a ride-hailing option and are asked to explain their reasoning in a free-text response.}
\end{figure*}

\subsection{Apparatus}\label{sec:apparatus}
We developed a custom web-based prototype of a ride-hailing service named \textit{RidePal} for this purpose. The server-side implementation was built in Python (version 3.13.6)\footnote{https://www.python.org/doc/versions/ – Accessed: 24/01/2026} using the Flask framework (version 3.1.1)\footnote{https://flask.palletsprojects.com/en/stable/ – Accessed: 24/01/2026} to manage server–client communication. The frontend and dynamic interface behavior were implemented in JavaScript. Participant decisions were recorded through a logging system that generated a separate CSV file for each user. These files captured all relevant decision data, including trip selections, timestamps, ratings, and free-text answers. The positions of the maps and the positions of the items on the choice interface (see \autoref{fig:study-interface}a and \autoref{fig:study-interface}b) were randomized. To create realistic routes, we used the trip distances described in~\autoref{subsubsec:independent_vars} and leveraged the OpenRouteService Directions API\footnote{https://openrouteservice.org/ - Accessed: 24/01/2026} to generate realistic routes that matched these distances. The API provided both the route geometries and the corresponding walking and driving durations. Map tiles for rendering these routes were obtained from the OpenStreetMap tile server (via Leaflet)\footnote{https://leafletjs.com/reference.html - Accessed: 24/01/2026}. To present participants with static but realistic route displays, we used Puppeteer (headless Chrome)\footnote{https://pptr.dev/ - Accessed: 24/01/2026} to render the Leaflet maps in the browser and save screenshot pairs for each trip, showing both the walking and the driving option. 


\subsection{Procedure}\label{sec:procedure}
\begin{figure*}[h]
    \centering
    \includegraphics[width=0.95\linewidth]{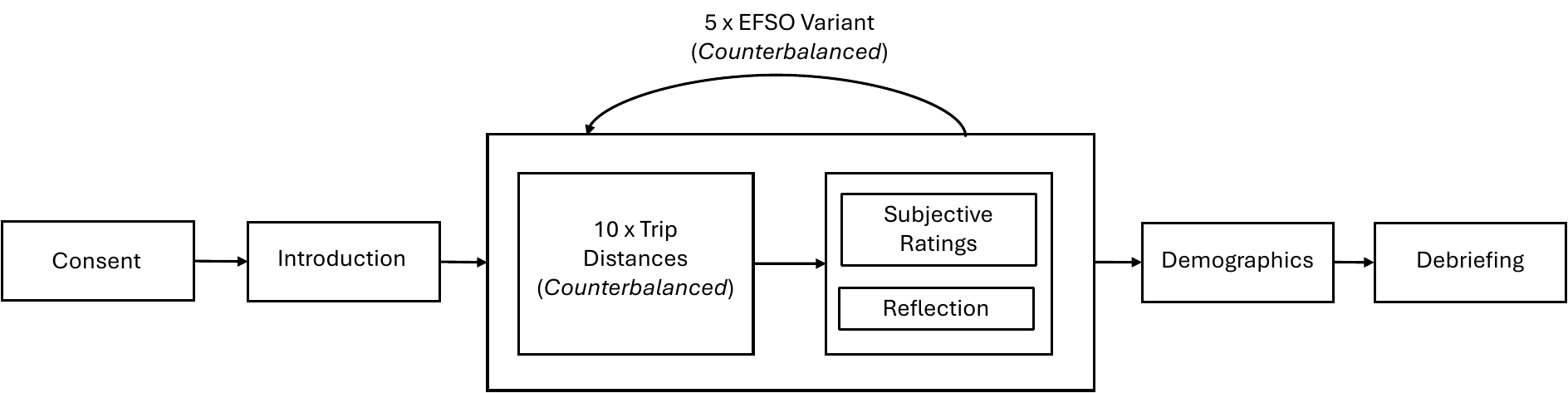}
    \caption{A visual process diagram showing a schematic overview of the study procedure as described in \autoref{sec:procedure}.}
    \label{fig:procedure}
    \Description{This figure outlines the study procedure. Participants first provided informed consent and read instructions. They then completed a simulated transportation choice task, making decisions between walking and ride-hailing across five experimental conditions, with ten trip decisions per condition and counterbalanced order. After each block, participants rated Likert-scale items and wrote a brief reflection on their choices. At the end of the task, they provided open feedback, answered demographic questions, and received a debriefing. The session took approximately 30 minutes, followed institutional ethics guidelines, and participants were compensated for their time.}
\end{figure*}

The experimental procedure is summarized below and visually depicted in \autoref{fig:procedure}. The study was listed on Prolific.com\footnote{https://www.prolific.com/ – Accessed 24/01/2026}, where participants could register. After enrollment, they were redirected to our university's LimeSurvey instance~\footnote{https://www.limesurvey.org – Accessed: 24/01/2026}, where they were assigned a unique participant identifier and provided informed consent. Participants then received an introduction that described the upcoming task, which was explained as a simulated transportation choice exercise involving a series of trip decisions. Once they confirmed that they had read the introductory information, they were redirected to the interactive task environment (see~\autoref{sec:apparatus}), which opened in a new browser tab.

Participants first received a description of the prototypical ride-hailing service, framed as comparable to services such as Uber or Lyft, along with information about the urban environment, which was described as walkable, equipped with sidewalks, and with no heavy traffic. They were then presented with the scenario description (see~\autoref{subsec:task}) outlining the travel context for all upcoming trips. After this introduction, participants completed ten trip decisions for each of the five EFSO conditions. The task was organized into blocks, with all ten trips for one EFSO variant completed before moving on to the next variant.

After each block, participants rated the single-item questions (see~\autoref{subsubsec:dep_var_moral_credit} and~\autoref{subsubsec:dep_var_perceived_impact}). They were then shown the trip distance at which they had first switched from walking to ride-hailing and were asked to provide an open-ended reflection on this decision (see~\autoref{subsubsec:dep_var_perceptions}). The reflection focused on the shortest trip distance where the switch occurred (see~\autoref{fig:study-interface}c), representing each participant’s individual threshold. Both the order of EFSO variants and the presentation of trip distances were counterbalanced using a balanced Latin square design~\cite{bradley1958complete}.

Upon completing decisions for all EFSO variants at each distance, participants were presented with a unique completion code that served as verification for returning to LimeSurvey. There, they were asked to provide demographic information and received a debriefing about the study’s purpose. Participants were compensated at a rate of £7.60 per hour, and the session lasted approximately 30 minutes. The study was conducted in accordance with institutional ethics and data protection guidelines. As no potential risks to participants were identified, a formal ethics review was not required under university policy.

\subsection{Participants}\label{subsec:sample}
In total, 89 participants were initially recruited via Prolific.com\footnote{https://www.prolific.com/ - Accessed: 24/01/2026}. As the studies that we used to ground the selected trip distances on were based on U.S. populations, we recruited participants from the U.S. After excluding data from participants who experienced technical difficulties or had otherwise unfinished response files, 75 remained for the analysis. The mean reported age was $M=36.8$ years ($SD = 9.47$). In terms of gender, 40 participants identified as female, 34 as male, and 1 as non-binary. In terms of educational backgrounds, 16 participants reported a high school degree, 38 held a bachelor’s degree, 11 a master’s degree, 4 a doctorate, 3 a professional degree (e.g., MD, DDS, DVM), and 3 reported another form of education. Participants' environmental attitude was measured using the New Environmental Paradigm~\cite{dunlap_new_2008} scale. Scores ranged from 2 to 5, with a mean of $M = 3.65$ ($SD = 0.75$), indicating a moderate to strong pro-environmental orientation among participants. In terms of mobility practices regarding ride-hailing, participants reported using, on average, approximately two different ride-hailing services ($M = 1.95$, $SD = 1.26$). 46 participants used both Uber and Lyft, 20 used only Uber, 6 used only Lyft, and 3 participants used other ride-hailing services than Uber and Lyft.
 
\subsection{Analysis}\label{subsec:analysis}
We examined the effect of each EFSO variant on ride-hailing choices using a logistic mixed-effects (LME) model implemented with the \texttt{lme4} package in R \cite{bates_fitting_2015}. Analyses were conducted in R version 4.5.1\footnote{https://cran.r-project.org/src/base/R-4/ - Accessed: 24/01/2026}, with all packages updated to their most recent versions. 

Because our research question centered on whether participants would choose to walk or to ride, the dependent variable was modeled as a binary outcome ($walk = 0,~ride = 1$). This operationalization captures the core behavioral trade-off of interest, which is substituting a zero-emission alternative (walking) for the less sustainable option of ride-hailing. Although participants always viewed three alternatives (walking and two ride-hailing options), the primary decision structure was first whether to walk or to ride, followed by which ride-hailing option to select.

We therefore conducted descriptive analyses for the type of ride-hailing service chosen but restricted inferential analyses to the binary choice between walking and ride-hailing, as this directly addressed our RQs and hypotheses (see \autoref{sec:method}). 

The LME model included EFSO variant, trip distance, and their interaction as fixed effects, with random intercepts and random slopes for distance specified by participant. An initial specification with additional random intercepts for condition resulted in singular boundary fit warnings, indicating overparameterization~\cite{bates_fitting_2015}. Removing this term allowed successful convergence ($AIC = 1994.0$, $BIC = 2074.9$, $\ell = -984.0$). The final model specification was: \[\text{Choice} \sim \text{EFSO Variant} \times \text{Distance} \; + \; (1 + \text{Distance} \mid \text{PID})\] To examine differences in how the EFSOs were perceived, we tested whether participants’ perceptions of environmental impact and perceived moral credit differed significantly between EFSO variants. Because Shapiro–Wilk tests indicated non-normality, non-parametric Friedman tests were applied, followed by Holm-corrected post-hoc tests for pairwise comparisons.

\section{Results}\label{sec:results}
\begin{figure*}[t]
    \centering
    \includegraphics[width=0.8\linewidth]{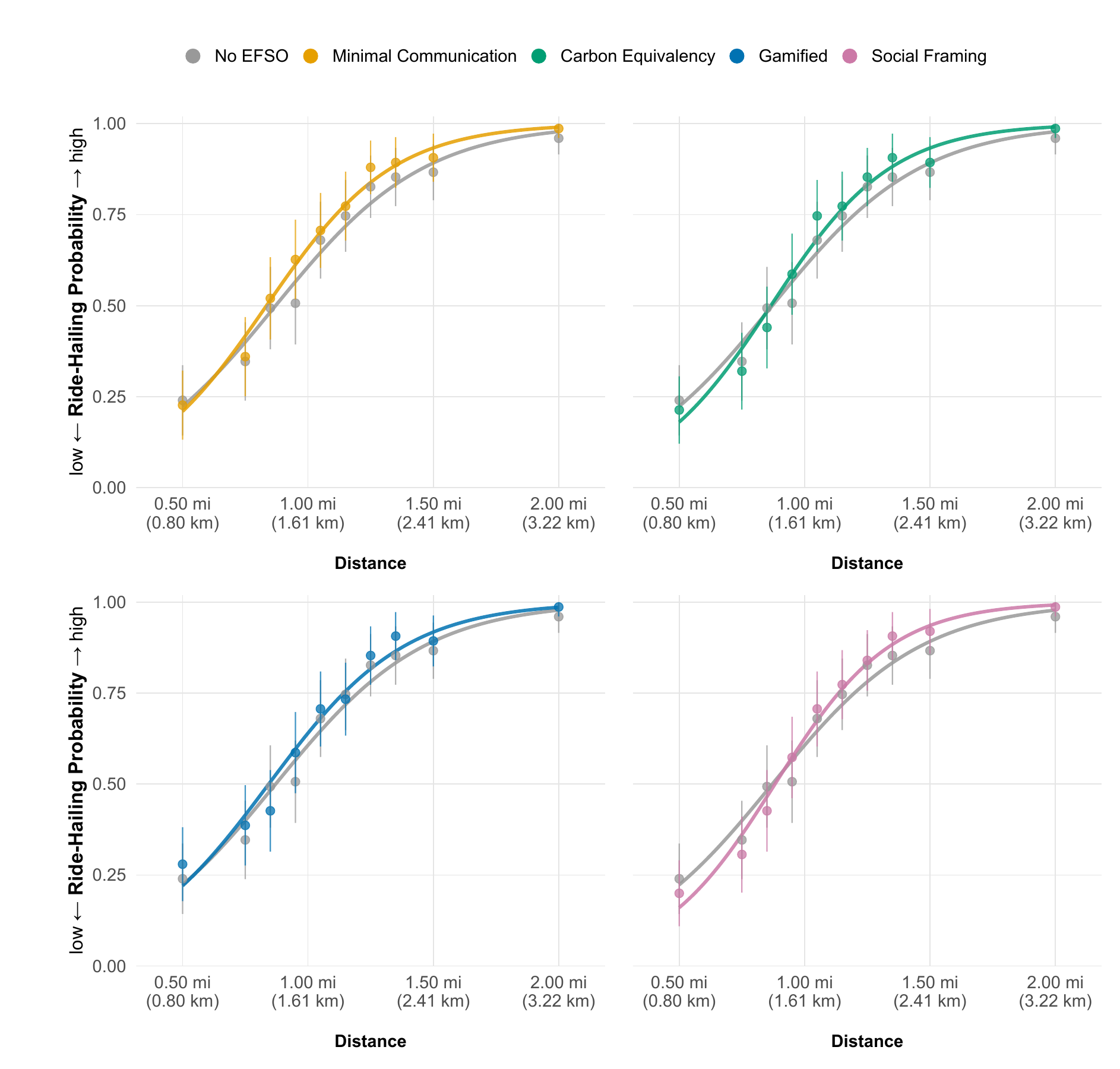}
    \caption{Figure showing four plots with observed probability of choosing ride-hailing over walking at each trip distance. Each plot shows the contrast between \baseline~ and either \mincom~(top-left), \coeq~(top-right), \gamified~(bottom-left), or \social~(bottom-right). Dots represent the mean probabilities, and error bars indicate 95\% confidence intervals. Distances are shown in miles and kilometers.}
    \label{fig:likelihood}
    \Description{The figure contains four separate plots showing how often participants chose ride-hailing instead of walking at different trip distances. Each plot compares the baseline condition (No EFSO), shown in gray, with one eco-friendly service option. The minimal-communication condition is shown in yellow, the carbon-equivalency condition in green, the gamified condition in blue, and the social-comparison condition in purple. Dots indicate the average probability of choosing ride-hailing at each distance, and the vertical bars represent 95 percent confidence intervals. Distances are labeled in both miles and kilometers. The plots show that the minimal-communication condition produces a small but nearly uniform increase in ride-hailing choices across all distances. The gamified condition closely overlaps with the No EFSO gray line, indicating almost no difference. For the carbon-equivalency and social-comparison conditions, the probability of choosing ride-hailing is lower than No EFSO for trips up to about one mile, but becomes higher than No EFSO for longer trips.}
\end{figure*}

\subsection{Transportation Choices}\label{subsec:transportation_choices}
\subsubsection{Descriptive Statistics}\label{subsubsec:res_desc_stats}
On average $M = 6.52$ ($SD = 2.74$) out of 10 possible rides for \baseline~ were taken, $M = 6.88$ ($SD = 2.54$) in \mincom, $M = 6.72$ ($SD = 2.52$) in \coeq, $M = 6.76$ ($SD = 2.72$) in \gamified, and $M = 6.64$ ($SD = 2.56$) in \social. 
$62.30\%$ of rides taken in \mincom~used the eco option, while $74.01\%$ of rides in \coeq~used the eco option. In \gamified~$68.05\%$ of rides used the eco option, and in \social~it was $68.67\%$. 
On average, participants switched from walking to ride-hailing at $M = 0.94$ ($SD = 0.38$) miles in \baseline~($M = 1.51km$, $SD = 0.61km$), $M = 0.91$ ($SD = 0.36$) miles in \mincom~($M = 1.46$, $SD = 0.58$) km, $M = 0.93$ miles ($SD = 0.36$) in \coeq~($M = 1.50km$, $SD = 0.58km$), $M = 0.92$ ($SD = 0.40$) miles in \gamified~($M = 1.48km$, $SD = 0.64km$), and $M = 0.94$ ($SD = 0.34$) miles in \social~($M = 1.51km$, $SD = 0.55km$).

\subsubsection{Fixed Effects of EFSOs on Transportation Choices}\label{subsubsec:res_average_effects}
\begin{figure}[t]
    \centering
    \includegraphics[width=1\linewidth]{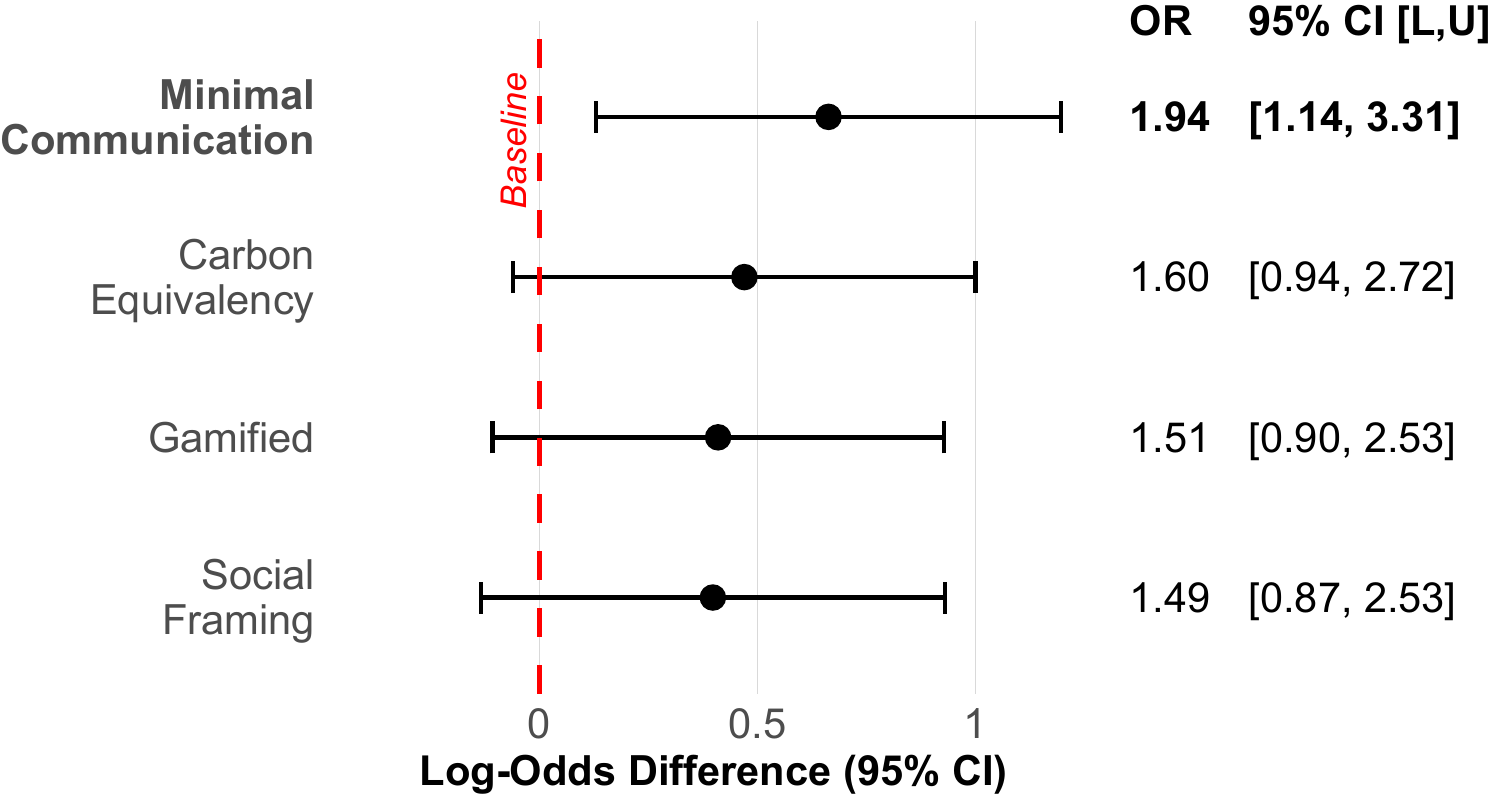}
    \caption{Figure showing the log-odds contrasts compared to \baseline. The dashed horizontal line indicates the no-difference threshold relative to \baseline, and error bars represent 95\% confidence intervals. The panel on the right displays the corresponding odds ratios (OR) along with their 95\% confidence intervals, with L denoting the lower bound and U the upper bound.}
    \label{fig:forest-plot}
    \Description{The figure presents log-odds contrasts for each eco-friendly service option compared to the baseline condition. Each EFSO has its own point estimate, accompanied by a horizontal error bar indicating the 95 percent confidence interval. A red vertical dashed line marks the threshold, indicating no difference relative to baseline. To the right of the log-odds plot, the corresponding odds ratios and their 95 percent confidence intervals are listed. The contrast between the minimal-communication EFSO and the no-EFSO baseline shows the strongest positive effect on ride-hailing choices. Its odds ratio is 1.94, with a confidence interval ranging from 1.14 to 3.31, indicating a notable increase in the likelihood of choosing ride-hailing under this condition.}
\end{figure}

Relative to \baseline, in \mincom~($\beta= 0.66$, $SE = 0.22$, $z = 3.05$, $p = .002$) and \coeq~($\beta = 0.47$, $SE = 0.22$, $z = 2.18$, $p = .029$) participants were significantly more likely to choose ride-hailing. The effects of \gamified~($\beta = 0.41$, $SE = 0.21$, $z = 1.94$, $p = .052$) and \social~($\beta = 0.40$, $SE = 0.22$, $z = 1.84$, $p = .066$) were not statistically significant. Contrasts with \baseline, and \mincom~as the reference level showed that, when averaged over the entire distance range with multiple comparisons correction, only the contrast of \mincom~and \baseline~remained statistically significant ($\beta = 0.66$, $SE = 0.22$, $z = 3.05$, $p = .019$, $OR = 1.94,~95\% CI~[1.14,~3.31]$). No statistically significant difference in the contrasts that used \mincom~as the reference level was found. Thus, on average, there was no reliable difference in ride-hailing uptake between presenting participants with an EFSO with no eco-feedback compared to when they were presented with EFSOs that did include eco-feedback (i.e.~\coeq, \gamified, \social).~\autoref{fig:forest-plot} illustrates the log-odds contrasts using \baseline~as the reference level.

\begin{figure}[b]
    \centering
    \includegraphics[width=1\linewidth]{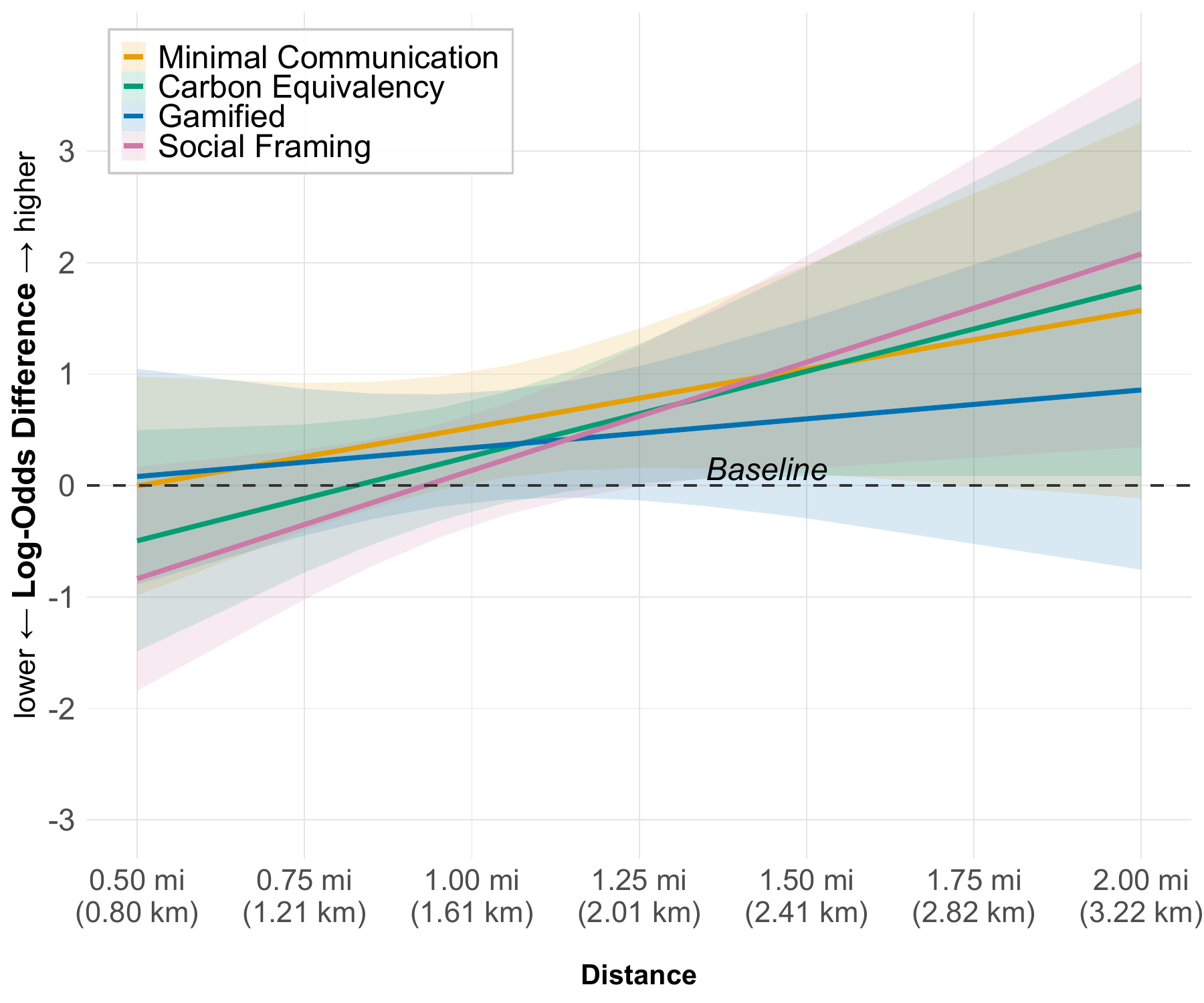}
    \caption{Figure showing the log-odds differences relative to \baseline~ across the full distance range. The dashed horizontal line marks the no-difference threshold, and the shaded ribbons depict the 95\% confidence intervals around the estimated contrasts. Distances are shown in miles and kilometers.}
    \label{fig:interaction_plot}
    \Description{The figure displays four colored lines showing how each eco-friendly service option differs from the baseline condition in log-odds of choosing ride-hailing across trip distances. The baseline reference is indicated by a dashed horizontal line marking the no-difference threshold. Each EFSO condition has its own line and matching shaded ribbon showing the 95 percent confidence interval. Yellow for the minimal-communication option, green for the carbon-equivalency option, blue for the gamified option, and purple for the social-comparison option. The vertical axis represents the log-odds difference relative to the baseline, with values above zero indicating higher ride-hailing likelihood and values below zero indicating lower likelihood. The horizontal axis shows trip distance from 0.5 to 2 miles.}
\end{figure}

As expected, trip distance was a significant predictor of ride-hailing choices ($\beta = 11.39$, $SE = 1.25$, $z = 9.13$, $p < .001$). Significant interactions effects of \textit{EFSO Variant} $\times$ \textit{Distance} were found for \coeq~($\beta = 1.52$, $SE = 0.68$, $z = 2.25$, $p = .025$) and \social~($\beta = 1.94$, $SE = 0.69$, $z = 2.82$, $p = .005$). Based on this, the rate of ride-hailing uptake increased significantly for these EFSO variants as trip distances grew longer. Interactions for \mincom~($\beta = 1.05$, $SE = 0.67$, $z = 1.57$, $p = .116$) and \gamified~($\beta = 0.52$, $SE = 0.64$, $z = 0.80$, $p = .421$) did not reach statistical significance. Interactions across the distance range are shown in~\autoref{fig:interaction_plot}.

\subsubsection{Effects of EFSOs Across Distances}\label{subsubsec:res_distance_effects}
To identify the distances at which ride-hailing adoption shifts due to EFSOs occurred, pairwise comparisons, adjusted for multiple comparisons. 
In \mincom~were significantly more likely to choose ride-hailing than those in \baseline~at $1.05mi~(\approx1.69km)$ ($OR = 1.78$, $p = .019$), $1.15mi~(\approx1.85km)$ ($OR = 1.97$, $p = .009$), $1.25mi~(\approx2.01km)$ ($OR = 2.19$, $p = .009$), $1.35mi~(\approx2.17km)$ ($OR = 2.43$, $p = .012$), and $1.50mi~(\approx2.41km)$ ($OR = 2.85$, $p = .024$).
Similar effects were observed for \coeq, which showed higher likelihood of ride-hailing compared to \baseline~at $1.25mi~(\approx2.01km)$ ($OR = 1.91$, $p = .045$), $1.35mi~(\approx2.17km)$ ($OR = 2.22$, $p = .032$), $1.50mi~(\approx2.41km)$ ($OR = 2.79$, $p = .029$), and $2.0mi~(\approx3.22km)$ ($OR = 5.96$, $p = .039$).
\social~differed from \baseline~at $1.35mi~(\approx2.17km)$ ($OR = 2.26$, $p = .029$), $1.50mi~(\approx2.41km)$ ($OR = 3.03$, $p = .017$), and $2.0mi~(\approx3.22km)$ ($OR = 7.99$, $p = .013$). No other effects were found using \baseline~as the reference level. Again, no other differences between \mincom~and the more information-rich interventions reached significance. In sum, at distances below $1.05mi$ ($\approx$20 minutes walking time), none of the choices that included EFSOs statistically significantly shifted the likelihood of ride-hailing adoption. 

\begin{figure*}[t]
    \centering
    \includegraphics[width=1\linewidth]{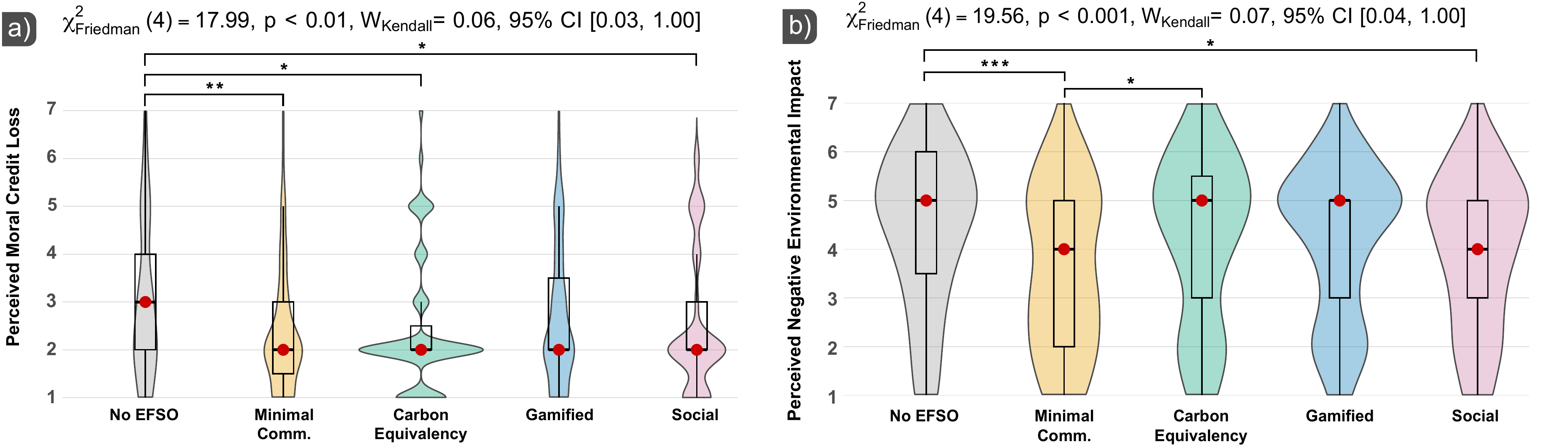}
    \caption{This figure shows the test statistic for the statistical comparisons in terms of perceived moral credit loss ratings in (a) and the perceived environmental burden compared to walking in (b). Asterisks denote significant differences ($*=p<0.05,~**=p<0.01,~*** = p<0.001$)}
    \label{fig:manipulation-check}
    \Description{This figure presents the test statistics for the comparisons of perceived moral credit loss (panel a) and perceived environmental burden relative to walking (panel b). In panel (a), baseline scores were significantly higher than those in the minimal-communication, carbon equivalency, and social comparison EFSO. In panel (b), perceived environmental burden was significantly higher in the baseline condition compared to the minimal communication and social comparison condition. The scores for minimal communication were also significantly lower than those for the EFSO that used carbon equivalencies. No other statistically significant differences were observed between conditions in perceived environmental burden when participants switched from walking to ride-hailing.}
\end{figure*}

\subsection{Subjective Ratings}\label{subsec:res_subjective_ratings}
\subsubsection{Perceived Moral Credit Loss}\label{subsubsec:res_moral_credit}
A non-parametric Friedman test found a significant effect of EFSO Variant on perceived moral credit loss ratings ($\chi^2(4) = 17.99$, $p < .01$, Kendall's $W = 0.06$). Holm-corrected post-hoc pairwise comparisons show that scores in \baseline~($Mdn = 3.00$) were significantly higher than in \mincom~($Mdn = 2.00$), $p_{adj} < 0.01$, $r = 0.52$, \coeq~($Mdn = 2.00$), $p_{adj} < .05$, $r = 0.42$, and \social~($Mdn = 2.00$), $p_{adj} < .05$, $r = 0.40$. As depicted in~\autoref{fig:manipulation-check}a, no other statistically significant differences were found.

\subsubsection{Perceived Negative Environmental Impact}\label{subsubsec:res_environmental}
A non-parametric Friedman test found a significant effect of the encountered EFSO variant on perceived negative environmental impact ratings ($\chi^2(4) = 19.56$,~$p < .001$,~Kendall's $W = 0.07$). Holm-corrected post-hoc pairwise comparisons show that for \baseline~($Mdn = 5.00$) perceived negative environmental impact was significantly higher than in \mincom~($Mdn = 4.00$), $p_{adj} < .001$, $r = 0.46$ and \social~($Mdn = 4.00$), $p_{adj} < .05$, $r = 0.41$. Scores in \mincom~($Mdn = 4.00$) were also significantly lower than in \coeq~($Mdn = 5.00$), $p_{adj} < .05$, $r = 0.47$. As depicted in~\autoref{fig:manipulation-check}b, no other statistically significant differences were found.

\subsection{Qualitative Data}\label{subsec:qualitative}
\subsubsection{Analysis}  
We analyzed the qualitative responses using reflexive thematic analysis following \citet{clarke2017gareth}. All responses were compiled into a single dataset containing participant IDs, the textual responses, and the corresponding EFSO variant. On average, responses were $M=21.22$ ($SD=17.08$) words long. Open coding was conducted in a single session using an inductive approach, with discussions taking place as they arose during the coding process. Points of discussion were primarily related to merging multiple codes with similar meaning and formulation. This resulted in 32 unique codes. In a second session, similar codes were then clustered into broader categories, which yielded the themes reported below, supported by excerpts from participant responses. The full codebook is provided in the supplementary materials.  

\subsubsection{Theme 1: Convenience and Comfort Dominate Decision-Making}
Across conditions, participants consistently framed their choices around convenience and physical comfort. Time efficiency emerged as the most salient factor, with many setting clear thresholds for when walking was acceptable and when it became unreasonably demanding, despite the study design not explicitly defining any time pressure in the trip scenarios. These thresholds were often expressed in minutes and served as personal cutoffs that guided decisions more strongly than cost or environmental impact.

\begin{quote}
``I switched at this distance because I didn't want to walk more than 15 minutes, and this [regular ride option] was the cheaper option.'' -P26
\end{quote}

Convenience was not only about time but also about minimizing physical effort and avoiding strain. Participants frequently referred to fatigue, the desire to remain comfortable, or to arrive in good condition for social encounters. Safety concerns were similarly folded into this logic, with participants citing the complexity of certain routes or the presence of traffic as reasons to favor a ride. These considerations highlight how the idea of ``comfort'' extended beyond bodily ease to include perceptions of security and peace of mind.

\begin{quote}
``This looked like the route went into a complicated area. I was worried that it wasn't actually safe to walk there, so I took a car at this distance.'' -P31
\end{quote}

Convenience often acted as the decisive trigger for switching from walking to ride-sharing. Participants described walking as acceptable only up to a certain threshold, after which the additional time or effort was framed as an unnecessary burden. Once that boundary was crossed, ride-hailing became the default option, with convenience overriding other considerations.

\subsubsection{Theme 2: EFSOs Enhance the Positive Perception of Convenience-Motivated Choices}
When participants encountered tensions between cost, time, and environmental values, EFSOs functioned as a mechanism for reconciling these competing priorities. They were often described as enabling a balance between personal convenience and broader responsibility. Although the scenarios were standardized with walkable infrastructure, no time pressure, safe routes, and casual meetings, participants still invoked these contextual factors to justify why ride-hailing appeared more sensible. The presence of an EFSO then served as an additional rationale, reinforcing the decision by presenting it as not only practical but also environmentally responsible.

\begin{quotation}
``My thought process was to prioritize time and convenience. The walking route is 15 minutes, which I consider to be a bit long when I'm in a hurry. The ride option is only 2 minutes and costs a reasonable amount, which is a great value for the time saved. While walking would be the most eco-friendly option, the gain of 2 x EcoPoints with the ride option made it feel like a slightly more environmentally conscious choice than a regular ride, while still prioritizing my need to get to the meeting quickly.'' -P3
\end{quotation}

In this way, EFSOs operated as moral and pragmatic compromises that helped participants legitimize their selections. This trade-off negotiation was also evident in relation to the price of the EFSO. In our study, the less carbon-intensive ride was always one dollar more expensive while being subtly (\mincom) or more clearly (\coeq, \social, \gamified) framed as less environmentally damaging. Participants' reflections when switching from walking to ride-hailing also leaned on the presence of EFSOs to mitigate feelings of guilt or conflict, positioning their decision as both convenient and ethically defensible.

\begin{quote}
  ``The ride saves me time and I can take the eco version so I don't feel quite as guilty.'' -P42  
\end{quote}

Environmental considerations, therefore, were not dismissed outright but were integrated into broader calculations shaped by time and affordability. Rather than radically altering decision-making, they provided a pathway to make convenient decisions feel more acceptable, embedding sustainability within existing rationales rather than overriding them.

\subsubsection{Theme 3: Views on the True Relevance of EFSOs Are Conflicting}
Beyond specific trip scenarios, participants reflected more broadly on the role of EFSOs in their everyday lives. Many regarded them as meaningful opportunities to contribute to sustainability in small but consistent ways. These participants valued the ability to integrate eco-friendly choices into routine practices without requiring major lifestyle changes, and some described a sense of pride or moral satisfaction when making such selections.

\begin{quote}
 ``The more people that partake in these eco-friendly options, the higher the impact. If we all pitch in together, we could be making significant changes to the environmental impact of CO2 emissions. These options mean a great deal to me because it gives me a chance to do my part to create a more eco-friendly lifestyle.'' -P12   
\end{quote}

Some participants expressed skepticism about the impact of EFSOs, pointing to cost premiums, doubts about providers’ claims, and concerns about greenwashing. They emphasized that while individual choices can contribute, meaningful progress depends on systemic changes led by companies and governments. From this perspective, when EFSOs were available, they were appreciated but regarded as insufficient when responsibility for sustainability appeared to be shifted primarily onto consumers.

\begin{quote}
    ``I perceive the impact of eco-friendly options in online services to be marginal, but meaningful. The actual carbon emissions saved by an individual choosing an "Eco Ride" or "green delivery" option is often a tiny fraction of their overall carbon footprint. However, their true value lies in their ability to raise awareness and normalize sustainable choices.'' -P55
\end{quote}

\section{Discussion}\label{sec:discussion}
%
\subsection{EFSOs and Behavioral Rebound in Ride-Hailing (RQ1 \& RQ2)}
\subsubsection{Introducing EFSOs can Facilitate Rebound Behaviors}
The present study examined whether the introduction of EFSOs influences ride-hailing uptake compared to a zero-emission alternative, such as walking, and how different eco-feedback metrics modulate this effect. The large random intercept and random slope variances in the LME model (see~\autoref{subsubsec:res_average_effects}) indicate that participants varied widely in their baseline tendency to choose ride-hailing over walking and in how strongly their choices changed as distance increased. Our Prolific recruitment did not restrict participants to specific U.S. regions or demographic groups, which likely contributed to the observed baseline variability. Across all conditions, participants selected ride-hailing in roughly two-thirds of trials and switched from walking to ride-hailing at $\approx0.9mi-1.0mi~(\approx1.44km-1.61km)$ (see~\autoref{subsubsec:res_desc_stats}). These behavioral outcomes of our experiment are consistent with estimates of reasonable walking distances reported in prior research. For instance, based on an analysis of a cross-sectional adult U.S. sample (N=3653),~\citet{watson_walking_2015} found that less than 43\% of their sample perceived walking distances beyond $1.0 mi$ as reasonable. This also coincides with \textit{Distance} being the main driver of mobility decisions in our experiment (see~\autoref{subsubsec:res_average_effects}).\\

In terms of how ride-hailing uptake shifted depending on the encountered EFSO variant, our analysis revealed an increased average odds of ride-hailing adoption when participants encountered \mincom~compared to \baseline. Averaged over the entire distance range, \coeq, \social, and \gamified~did not significantly shift ride-hailing adoption odds. However, effects for some EFSOs varied significantly depending on trip length, with \coeq~and \social~making ride-hailing uptake more likely as trips grew longer, even if choices at shorter distances did not initially differ from \baseline.

Distance-specific analyses further showed that no statistically significant shift in ride-hailing adoption occurred below $1.05mi~(\approx1.68km)$. Qualitative responses frequently indicated that participants viewed 15–20 minutes of walking (with 20 minutes approximately corresponding to $1.0mi~(\approx1.61km)$ according to the OpenRouteService Directions API\footnote{https://openrouteservice.org/— Accessed: 24/01/2026}) as a personal threshold for choosing to walk. Together, these patterns suggest that detectable shifts in ride-hailing adoption primarily occur when trips approach distances at which the effort of walking is approaching a personal threshold that can be balanced by choosing the ride-hailing alternative.

Based on the above-mentioned outcomes, there is \textbf{partial support for H1}. With the exception of \gamified, EFSOs influenced ride-hailing uptake at distances where participants were close to a natural point of uncertainty between walking and taking a ride (\mincom), or walking was already seen as not feasible (\coeq, \social). This suggests that within the scope of our experiment \textbf{EFSOs were able to facilitate rebound behaviors primarily when the underlying acceptable compromise between convenience and effort is already finely balanced (i.e., when trip distance extends beyond what is considered ``reasonable walking distance'')}. The shifts caused by \mincom, \coeq, and \social, while detectable statistically, were modest relative to the dominant effect of the variable \textit{Distance} itself.

While our data indicate that EFSOs facilitate behavioral patterns consistent with direct rebound effects, it is important to note that our study did not measure rebound effects in a strict sense. Rebound effects describe the reduction in expected environmental benefits once offsetting behavioral responses are taken into account~\cite{vivanco2016}. Our results, therefore, provide evidence that EFSOs enable behavioral tendencies that contribute to such effects, rather than demonstrating the rebound mechanism directly. These insights align with prior work showing that seemingly pro-environmental features or choices can inadvertently promote less sustainable behavioral patterns as an unintended consequence (e.g.,~\cite{urban_pro-environmental_2023, dutschke_moral_2018, truelove2014positive}).

\subsubsection{Eco-Feedback Does Not Amplify Rebound Behavior}
When directly comparing the results for the richer eco-feedback variants (\coeq, \gamified, \social) to \mincom, no statistically significant differences in ride-hailing adoption were observed. As such, \textbf{the data do not support H2}. Participants also rated switching from walking to ride-hailing when using \mincom~as less environmentally harmful than \baseline~or \coeq~and more morally desirable than switching from walking to ride-hailing using \baseline~(see~\autoref{sec:results}). Unlike \mincom, the richer feedback formats supplied concrete environmental information, such as \co~quantities or contextual analogies. Prior work suggests that such information increases environmental awareness~\cite{froehlich2010} and can provide relatable cues about environmental impacts~\cite{mohanty2023}. Based on this, one explanation for how \mincom~was interpreted by participants could be that \mincom~allowed for more unconstrained user interpretations of the ``eco-option''.\\

Regarding perceived moral credit loss, switching from walking to ride-hailing with \mincom, \coeq, and \social was rated as less morally costly than \baseline, whereas \gamified~did not differ significantly from \baseline. These mixed outcomes provide only \textbf{partial support for H3}, but viewed together with the qualitative responses (see~\autoref{subsec:qualitative}), they also indicate that EFSO designs differ in how permissive or consequential they make choosing an EFSO to users. Participants’ qualitative responses indicated that the meaning of the point-based framing in the \gamified~variant was not always understood as being connected to environmental implications, as no explanation accompanied the points. This lack of clarity may account for why \gamified~did not differ significantly in perceived moral credit loss, perceived negative environmental impact, or observed behavioral shifts. Comparable patterns were reported by \citet{mohanty2023}, who found that participants often inferred that point-based EFSOs must confer some benefit but were unsure what the benefit represented or whether it was genuinely pro-environmental. In terms of moral credit, our experiment did not test perceived moral credit loss as a direct mediator of choice behavior. The measures we included were instead used as contextual indicators of how participants understood the EFSO framings, rather than evidence of moral-psychological processes being the main driver of decisions. However, when considered alongside the qualitative reflections, many of which referenced justification, guilt reduction, or balancing convenience with environmental considerations, the results point to the relevance of moral-psychological reasoning in how users made sense of the EFSOs. Considering prior work on moral-psychological mechanisms underlying rebound effects (e.g.,~\cite{reimers2022moral, dutschke_moral_2018, sorrell_limits_2020, santarius_investigating_2016, santarius_how_2018_psychological_mechs}), information presentation can influence how impactful or relevant a decision with environmental consequences is perceived to be. While previous studies have primarily examined eco-feedback as a persuasive tool intended to support reflective decision-making~\cite{disalvoMapping2010, bremerReview2022, zeqiri2024}, our findings indicate that \textbf{eco-feedback and environmental impact communication should likely also be evaluated in terms of how they shape behavioral responses once an eco-friendly option becomes available and their adoption is increased.}\\

\subsection{User Rationales in the Presence of EFSOs (RQ3)}
A key aim of this study was to examine how participants rationalize decisions in the presence of EFSOs in order to better understand reasoning processes that could facilitate rebound. Participants frequently emphasized convenience and comfort as decisive thresholds for switching from walking to ride-hailing. Once these thresholds were crossed, the presence of an EFSO offered an additional layer of justification, allowing participants to frame their decision as a personally acceptable trade-off between convenience and environmental concern. 

The \mincom~condition in particular appeared to create room for interpretations of environmental benefits, enabling participants to present their choice as both practical and responsible. Richer feedback framings shaped this interpretive space by grounding the impact in metrics such as \co equivalencies, social comparisons, or gamified point systems. Consistent with the quantitative results, qualitative responses suggest that EFSOs did not fundamentally shift the convenience-dominated nature of decision-making and, in some cases, even introduced feelings of guilt. In practical terms, this aligns with rebound effect conceptualizations by~\citet{dutschke_moral_2018} and~\citet{reimers2022moral} and suggests that moral-psychological rationalizations should not be viewed in isolation as drivers of rebound but rather as one dimension that interacts with dominant priorities such as time efficiency and physical effort. Although our experimental task explicitly removed time pressure and guaranteed safe walking conditions (e.g., by describing the existence of sidewalks and regular traffic), several participants nonetheless expressed considerations such as arriving late if they walked or perceived certain routes as unsafe. These reactions suggest that participants drew on familiar mobility habits and personal narratives rather than limiting themselves to the conditions presented in the task. Trade-offs between convenience and environmental concern were not always described as explicit or deliberate considerations. Instead, participants reported that EFSOs made convenience-driven choices feel more acceptable without requiring much reflection. This aligns with the idea that individuals seek to maintain a coherent self-perception that is consistent with their actions~\cite{festinger1962cognitive}. The unintended legitimizing effects of EFSOs may therefore operate in subtle ways, rather than prompting a conscious evaluation of trade-offs. 

Overall, participants expressed divergent views on the broader significance of EFSOs. Some regarded them as meaningful opportunities to contribute to sustainability in everyday life, whereas others voiced skepticism, highlighting concerns about greenwashing and the limited impact of individual actions. Such diversity is consistent with recent work by~\citet{zeqiri2024}, who examined user perspectives on energy-saving features in household appliances. They found that while many users employ such features to balance environmental values with convenience or cost-effectiveness, others dismissed them as primarily ``marketing''. We suggest that EFSOs in online services may be perceived along a similar spectrum, ranging from meaningful contributions to sustainability to superficial gestures that shift responsibility onto consumers.  


\subsection{Practical Implications}  

\subsubsection{EFSOs Should Remain Available}  
While our results suggest that introducing EFSOs in this study shifted some user decisions toward less sustainable, comfort-oriented behaviors, we do not suggest that EFSOs should be avoided. Qualitative responses indicate that they were frequently regarded as valuable opportunities to contribute to sustainability in everyday life, and their availability can help normalize pro-environmental choices. Ensuring accessibility and visibility, therefore, remains important so that EFSOs continue to provide simple entry points for sustainable behavior. At the same time, the potential for unintended behavioral responses should be acknowledged, and integration strategies should be adapted accordingly based on the implications outlined below.  
  
\subsubsection{Communicate the True Magnitude of Benefits When Introducing EFSOs}  
Based on our findings, relying on users to estimate the impact of EFSOs from only minimal indicators is not advisable, as this leaves room for generous interpretations of the environmental benefits connected to EFSOs. Participants' choices during exposure to \mincom, for example, often described the eco-ride option as both convenient and environmentally preferable, despite the absence of information about its actual impact. If EFSOs are integrated, they should be accompanied by information that conveys the scale of their environmental benefits. Although convenience and time efficiency were the primary drivers of decisions, presenting transparent information about the realistic magnitude of carbon savings, for example, through absolute values or contextual equivalencies, can help reduce misperceptions.  
    
\subsubsection{Address Potential Moral-Balancing Dynamics in the Design Process}  
Our qualitative findings indicate that some participants used the presence of an EFSO to frame ride-hailing as an acceptable choice, suggesting reasoning patterns that resemble moral balancing. Additionally, the reflections indicate that users may integrate environmental justifications alongside convenience considerations. Designers could take such tendencies into account during the evaluation stage of the design process by systematically examining how environmental labels or cues reshape perceived moral trade-offs and whether they inadvertently legitimize increased consumption. This may involve testing prototypes in scenarios where users must navigate competing motivations, assessing whether environmental framing alters the threshold for choosing higher-emission options, and identifying conditions under which eco-features are used as justifications rather than supports for genuinely lower-impact behavior.  

\subsubsection{Moral-Psychological Pathways Should Be Considered as a Lens for Understanding Rebound Effects}  
Prior work in HCI has largely conceptualized rebound effects through economic mechanisms (e.g.,~\cite{bremerRebound2024}). While addressing economic incentives via interface design remains challenging, our findings align with environmental psychology research (e.g.,~\cite{reimers2022moral, dutschke_moral_2018, santarius_how_2018_psychological_mechs}) that highlights the role of moral-psychological pathways in shaping how individuals interpret environmentally relevant actions. The ascription of ``moral credit'' is closely tied to how people perceive their own behavior~\cite{merritt2010moral}, and our qualitative results show that participants sometimes used the presence of an EFSO to frame convenience-oriented ride-hailing choices as personally acceptable. Eco-feedback and eco-visualization techniques are well established in HCI~\cite{bremerDataCenters2025, disalvoMapping2010} and could offer a potential design space for influencing such interpretations. Recognizing these dynamics provides the HCI community with one of several complementary lenses for investigating why rebound effects occur when pro-environmental features or technologies are introduced.

\section{Limitations}\label{sec:limitations}
In the following sections, we acknowledge several limitations of our experiment, discuss their impact on the interpretation of our findings, and outline avenues for further research.

As described in \autoref{sec:method}, we selected an online experimental design to isolate how EFSOs influence ride-hailing decisions under conditions where contextual factors can be controlled. We standardized trip purpose, time availability, and route characteristics through the wording in our introduction and scenario description. This made it possible to attribute observed behavioral differences more directly to our experimental manipulations, which is a requirement for internal validity. However, real-world transportation decisions unfold in more dynamic settings. Contextual considerations vary from trip to trip and often interact in unpredictable ways. As a result, the magnitude of the effects may have been impacted by our wording and could differ in real-world settings. Similarly, the ratings regarding moral credit and perceived environmental burden were framed negatively, which could have biased participants to express rationales involving guilt more frequently. 

Another limitation is that our choice task simplified the structure of ride-hailing decisions. In the experiment, participants viewed all travel options simultaneously. Outside the lab, decisions can initially be sequential. Individuals first decide whether to use ride-hailing at all and then choose between vehicle types once inside the app. This simplification does not fully capture the temporal dynamics of naturalistic decision-making, particularly among users who may not yet know that an EFSO will be available. As such, it might have biased participants' choices in our experiment. Nevertheless, over time, as EFSOs become more common and predictable on ride-hailing platforms, users may begin to anticipate their availability earlier in their trip planning, for example, by allowing less time to walk or assuming that a lower-impact ride will be offered. Such anticipatory effects align with the rebound behaviors observed in our simultaneous-choice design, although this remains an empirical question that requires field-based verification in future work. 

Our sample consisted of only U.S. participants to maintain consistency with the U.S.-based investigations of reasonable walking distance perceptions that informed our experiment (\cite{watson_walking_2015, yang2012walking}). Although we did not restrict demographic groups during recruitment, the U.S.-based sample limits generalizability to other cultural contexts. We collected demographic information regarding ride-hailing services participants personally use, but did not collect data on walking preferences or physical limitations. Because each participant saw all EFSO variants at all distances, our design captured how EFSOs shifted relative preferences within the same individual, which reduces, but does not eliminate, the influence of individual walking preferences on the estimated effects. We acknowledge that habitual mobility patterns and physical constraints could shape absolute walking rates. Our scenario did not explicitly standardize climate factors, which might have introduced uncontrolled variation, given that participants were recruited from across the U.S.

The within-subjects design we employed introduces its own constraints as participants were exposed to all combinations of the variables \textit{EFSO Variant} and \textit{Distance}. It cannot be guaranteed that direct comparisons between EFSO variants did not additionally shape decisions in subtle ways. For instance, biases such as social desirability bias~\cite{nederhof1985methods} may have been triggered once participants became aware of the study's pro-environmental themes. To mitigate this, we fully counterbalanced the order in which participants encountered the different EFSO variants. We argue that, overall, this controlled foundation is essential for identifying which mechanisms merit closer examination in real-world studies or in other contexts, such as rural transportation or applications like travel booking. To gain additional confidence in the real-world applicability of our results, future work should reproduce the presented study design as a between-subjects design and test the effects in real-world studies.

Our experimental setup included walking as a zero-emission baseline to establish a clear behavioral contrast for identifying rebound-like responses. This contrast may not directly map to other service domains such as logistics or travel booking, where users typically cannot choose a truly zero-emission option. We view this as a methodological rather than a conceptual constraint. Rebound effects do not necessarily require the presence of a zero-emission option. They reflect increased consumption or reduced restraint following an eco-labeled choice. In domains without a zero-emission option, rebound may instead manifest as a greater willingness to choose higher-impact options or as an increase in consumption frequency. This changes how studies must be designed to detect rebound effects, but not whether such effects can occur. Studies in such contexts would require designs that track shifts in consumption thresholds or escalation patterns rather than substitution away from a zero-emission alternative.

\section{Conclusion}\label{sec:conclusion}
This paper examined whether the introduction of EFSOs in a transportation context can facilitate behavioral rebound and how different eco-feedback concepts shape this dynamic. In an online within-subjects experiment (N=75), participants chose between walking and ride-hailing across multiple scenarios. Our findings indicate that the presence of an EFSO increased the likelihood of substituting ride-hailing for walking at feasible walking distances, with minimal feedback producing the most pronounced shifts. Eco-feedback concepts, such as communicating \co values with equivalencies, social framings, increased ride-hailing choices compared to a control condition without EFSOs, but did not significantly differ from the minimal feedback condition. An analysis of user rationales showed that participants primarily prioritized convenience and time efficiency, and often used EFSOs to frame these choices as acceptable compromises, even when this meant substituting a zero-emission option like walking. Importantly, participants did not necessarily describe consciously balancing guilt or moral credit, suggesting that such processes may occur more subtly. Based on these findings, we recommend that EFSOs remain available, as they represent motivating examples of easily realizable pro-environmental behaviors. However, they should be designed to limit interpretative flexibility by clearly communicating their actual benefits, which eco-feedback concepts may be well-suited to do. We further recommend that designers consider how moral psychological pathways, including the tendency to align convenient choices with perceived environmental justification, may shape decision-making. Users naturally balance multiple goals, and particular framings can unintentionally shift this balance. Evaluating designs with these dynamics in mind can help ensure that eco-framed features do not unintentionally encourage increased consumption when this is not the intended outcome.

\section*{Open Science}
The source code for our choice task and all necessary scripts to reproduce the materials and task environment are available at https://github.com/az16/efso-task. 
\begin{acks}
We thank our study participants for their participation. We further thank the reviewers for their valuable feedback and Jan Ole Rixen for early feedback on the concept. Grammarly and GPT-5, were used for identifying and improving spelling, grammar, punctuation, and clarity, and for suggesting paraphrasing, without altering the original content or meaning. No AI-generated text was used verbatim. This work was supported by the Landesgraduiertenförderung (LGFG) Scholarship for PhD students.
\end{acks}

\bibliographystyle{ACM-Reference-Format}
\bibliography{sample-base}


\end{document}